\documentclass[epj]{svjour}
\usepackage{graphics}

\usepackage{graphicx,epsfig,wrapfig}

\usepackage{amsmath}
\usepackage{cite}
\usepackage{color}
\usepackage{latexsym}
\usepackage{amssymb}
\usepackage{dsfont}
\usepackage{multirow}
\usepackage{bm}
\usepackage[normalem]{ulem}
\newcommand{\be}{\begin{equation}}
\newcommand{\ee}{\end{equation}}
\newcommand{\ba}{\begin{eqnarray}}
\newcommand{\ea}{\end{eqnarray}}

\newcommand{\di}{\mathrm{d}}
\newcommand{\ud}{\mathrm{d}}
\newcommand{\uvec}[1]{\boldsymbol{#1}}

\newcommand{\uS}{\boldsymbol{S}}
\newcommand{\us}{\boldsymbol{s}}

\newcommand{\usigma}{\boldsymbol{\sigma}}
\newcommand{\uk}{\boldsymbol{k}}

\begin{document}
\hyphenation{Wig-ner}
\hyphenation{im-pact-pa-ra-me-ter}
\hyphenation{spa-ce}
\allowdisplaybreaks[2]
\title{Modelling 
the nucleon structure}

\author{M. Burkardt\inst{1} \and B. Pasquini\inst{2,}\inst{3}% etc
}                     % Do not remove
\institute{Department of Physics, New Mexico State, University, Las Cruces, New Mexico 88003, USA \and Dipartimento di Fisica, Universit\`a degli Studi di Pavia, Pavia 27100, Italy \and Istituto Nazionale di Fisica Nucleare, Pavia, Italy }
\abstract{
 We review the status of our understanding of nucleon  structure based on the modelling of different kinds of parton distributions.
We use the concept of generalized transverse momentum dependent parton distributions and Wigner distributions, which combine the features of transverse-momentum dependent parton distributions and generalized parton distributions. 
We revisit various quark models which account for different  aspects of these parton distributions.
We then identify  applications of these distributions to gain a simple interpretation of key properties of the  quark and gluon dynamics in the nucleon.
 % end of PACS codes
} %end of abstract
%\and
\maketitle

\section{Introduction}
\label{intro}
The nucleon as a strongly interacting many-body system of quarks and gluons offers such a rich phenomenology that 
models are crucial tools to unravel the many facets of its nonperturbative structure.
Although models oversimplify the complexity of QCD dynamics and are constructed to mimic certain selected aspects
of the underlying theory, they are almost unavoidable  when studying the partonic structure of the nucleon and often turned out to be crucial to 
open the way to many theoretical advances.

Recently, a new type of distribution functions, known as generalized transverse momentum dependent parton distributions (GTMDs), has emerged 
as key quantities to study  the parton structure of the nucleon~\cite{Meissner:2009ww,Meissner:2008ay,Lorce:2013pza}.
They parametrize the unintegrated
off-diagonal quark-quark correlator, depending on the three-momentum $\uvec{k}$ of the quark and on the four-momentum $\Delta$ which is transferred by the probe to the hadron.
They have a direct connection with the Wigner distributions of the parton-hadron system, which represent the quantum-mechanical
analogues of classical phase-space distributions. Wigner
distributions provide five-dimensional (two position and three momentum coordinates) images
of the nucleon as seen in the infinite-momentum frame~\cite{Ji:2003ak,Belitsky:2003nz,Lorce:2011kd}. As such they contain the full
correlations between the quark transverse position and three-momentum. 

\begin{figure}[t!]
\begin{center}
\resizebox{0.4\textwidth}{!}{%
  \includegraphics{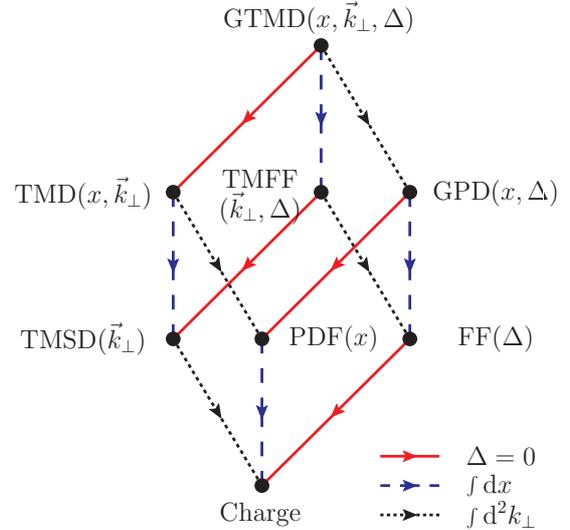}%,  width=0.5\columnwidth}
  }
\end{center}
\caption{\footnotesize{Representation of the projections of the GTMDs into 
parton distributions and form factors. }}
\label{fig1}
\end{figure}

In specific limits or after specific integrations of GTMDs,
 one can build up a natural interpretation of measured observables
known as generalized parton distributions (GPDs) and transverse momentum-dependent parton distributions (TMDs). 
Further limits/integrations reduce them to collinear
parton distribution functions (PDFs) and form factors (FFs) (see Fig~\ref{fig1} for a pictorial representation of the different links to GTMDs~\cite{Lorce:2011dv}).  

The aim of this work is to review the most recent developments in modelling the GTMDs, Wigner distributions, GPDs and TMDs, discussing the complementary and novel aspects encoded in these  distributions.
In sect.~\ref{sec:1} we  will focus on the GTMDs. 
As unifying formalism for modelling such functions, we will adopt the language of light-front wave functions (LFWFs), 
providing a representation of nucleon GTMDs which can be easily adopted in
many model calculations. 
In sect.~\ref{sec:2} we will discuss the derivation of the Wigner distributions. Although these phase-space distributions do not have a simple probabilistic interpretation, they entail a rich physics  
and we will discuss certain situations where a semiclassical interpretation is still possible.
\\
An important aspect that makes  Wigner
distribution particularly attractive is the possibility to calculate the expectation
value of any single-particle physical observable from its phase-space
average with the Wigner distribution as weighting factor.
In particular, the quark orbital angular momentum (OAM) can be obtained from the Wigner distribution for unpolarized
quarks in the longitudinally polarized nucleon~\cite{Lorce:2011ni,Hatta:2011ku}.
Recent developments~\cite{Burkardt:2012sd} have  shown that,
 depending on the choice of the Wilson path which makes the GTMD correlator gauge invariant, one can recover different  definitions of the OAM,  as discussed in 
sect.~\ref{sec:3}.
In sect.~\ref{sec:4} we focus the discussion on TMDs, giving an overview of different quark model calculations.
In sect.~\ref{sec:5} we present the GPDs, both in momentum and impact-parameter space, providing intuitive interpretations to relate the transverse distortions in impact-parameter space with the spin and orbital angular momentum structure of the nucleon.
Although a direct link between GPDs and TMDs does not exist, we discuss in sect.~\ref{sec:6} how one can relate, with a few simplifying assumptions,  the transverse shift in impact-parameter space described by the GPDs to the asymmetry in momentum space observed in semi-inclusive deep-inelastic scattering (DIS) and described by the TMDs.
In sect.~\ref{sec:7}, we extend the discussion to higher-twist PDFs. At variance with the leading-twist distributions, they do not have a simple partonic interpretation. Nevertheless,  it is possible to provide an intuitive explanation which allows one to connect the quark-gluon correlations described by these functions with  the transverse force generating momentum asymmetry in semi-inclusive deep-inelastic scattering (DIS). Our conclusions and perspectives are drawn in the final section.

\section{Generalized Transverse Momentum Dependent Parton distributions}
\label{sec:1}

The GTMDs are defined through the quark-quark correlator for a nucleon 
\ba
&&W^{[\Gamma]}_{\Lambda'\Lambda}(P, x,\uvec k_\perp,\Delta,{\cal W})\nonumber\\
&&=\int \frac{\di z^-\di^2z_\perp}{2(2\pi)^3}
\,e^{i  k\cdot z}\,\langle p',\Lambda'|\overline\psi(-\tfrac{z}{2})\Gamma\,\mathcal W\psi(\tfrac{z}{2})|p,\Lambda\rangle\rvert_{z^{+}=0},\nonumber\\
&&
\label{eq:GTMD}
\ea
which depends on the initial (final) hadron light-front helicity $\Lambda$ ($\Lambda'$), the  choice of the Wilson line ${\cal W}$ and the following independent four-vectors: 
\begin{itemize}
\item the average nucleon four-momentum $P=\tfrac{1}{2}(p'+p)$;
\item the average quark four-momentum $k$, with $k^+=x P^+$;
\item the four-momentum transfer $\Delta=p'-p$.
\end{itemize}
Note that we consider the light-cone plus-momenta $p^+$ and $p'^+$ to be large\footnote{Light-front coordinates of a generic four-vector $a=[a^+,a^-,\uvec a_\perp]$ are defined by $a^+=\tfrac{1}{\sqrt{2}}(a^0+a^3)$, $a^-=\tfrac{1}{\sqrt{2}}(a^0-a^3)$, and $\uvec a_\perp=(a^1,a^2)$.}.
The object $\Gamma$ in eq.~\eqref{eq:GTMD} stands  for any element of the basis $\{\mathds 1,\gamma_5,\gamma^\mu,\gamma^\mu\gamma_5,i\sigma^{\mu\nu}\gamma_5\}$ in Dirac space. 
The Wilson contour  $\mathcal W\equiv\mathcal W(-\tfrac{z}{2},\tfrac{z}{2}|\eta n_-)$ ensures the color gauge invariance of the correlators,
connecting the points $-\tfrac{z}{2}$ and $\tfrac{z}{2}$ \emph{via} the intermediate points $-\tfrac{z}{2}+\eta\infty n_-$ and $\tfrac{z}{2}+\eta\infty n_-$, with the four-vector $n_-=\tfrac{1}{\sqrt{2}}(1,0,0,-1)$ and 
the parameter $\eta=\pm$ indicating whether the Wilson contour is future-pointing or past-pointing.

Here we will discuss the formalism to calculate the GTMDs in quark models, without taking into account explicit gauge-field degrees of freedom, and, in particular, the contribution from the Wilson lines.  
\\
We assume that
the nucleon is described as an ensemble of non-interacting partons, which can be considered as a generic 
prototype for parton-model approaches~\cite{Jackson:1989ph,Roberts:1996ub,Efremov:2009ze,D'Alesio:2009kv,Bourrely:2010ng,Anselmino:2011ch}.
We will work in light-cone quantization in the $A^+=0$ gauge, which is 
the most natural framework to evaluate
the correlator~\eqref{eq:GTMD} at fixed light-cone time $z^+=0$.
The Fourier expansion of the quark field reads 
\ba
  &&\hspace{-0.7 truecm} \psi(z^-, \bm{z}_\perp)=\int\frac{{\rm d}k^+{\rm d}^2k_\perp}{2k^+(2\pi)^3}\,
   \Theta(k^+)\nonumber\\
   &&\hspace{-0.7 truecm}\times \sum_\lambda\left\{
   b^q_\lambda(\tilde k)u_\lambda(\tilde k)\, e^{-ik\cdot z}
   +d^{q\dagger}_\lambda(\tilde k)v_\lambda(\tilde k)\, 
   e^{ik\cdot z}\right\}\rvert_{z^{+}=0},\label{fourier}  
    \ea
    where $b^q$ ($d^{q\dagger}$) is  the annihilation (creation) operator of the quark (antiquark) field
and $u(\tilde k,\lambda)$ ($v(\tilde k,\lambda)$) is the free light-front Dirac spinor (antispinor).
Furthermore, $\lambda$ is the light-front helicity of the partons and $\tilde k$ denotes the light-front momentum variable $\tilde k=(k^+,\boldsymbol{k}_{\perp})$.
Using~\eqref{fourier}, and restricting ourselves to the quark contribution (and therefore to the region $\xi< x<1$, with $\xi=-\Delta^+/(2P^+)$),
  the operator in the correlator~\eqref{eq:GTMD} reads
\begin{eqnarray}\label{operator}
   &&\hspace{-1 truecm}\int \frac{\di z^-\di^2z_\perp}{2(2\pi)^3} \, 
   e^{i  k \cdot z} \, \overline{\psi}(-\tfrac{z}{2})\Gamma\psi(\tfrac{z}{2}) |_{z^+=0}\nonumber\\
    &&\hspace{-1 truecm}=2\int\frac{{\rm d}k^+{\rm d}^2k_\perp}{2k^+(2\pi)^3} \Theta( k^+)    
   \int\frac{{\rm d}k'^+{\rm d}^2k'_\perp}{2k'^+(2\pi)^3}\,\Theta( k'^+)\nonumber\\
   &&\hspace{-1 truecm}\times\delta(2 x P^+- k'^+- k^+)\delta^2(2 k_\perp - k'_\perp- k_\perp)\, \nonumber\\
   &&\hspace{-1 truecm}\times\sum_{\lambda,\lambda'}\overline u_{\lambda'}(\tilde{k}')\Gamma 
   u_\lambda(\tilde{k})\, b_{\lambda'}^{q\dagger}(\tilde{k}')b^{q}_{\lambda}(\tilde{k}).
   \end{eqnarray} 
     From eq.~\eqref{operator}, one easily recognizes the expression for the 
 full one-quark density operator in momentum  and light-front helicity space.
\\
 The nucleon state in \eqref{eq:GTMD} can be expanded in 
the light-front Fock-space  in terms of free
on mass-shell parton states, with the essential QCD bound-state information encoded in
the LFWF, i.e.
\begin{eqnarray}
\label{LFWF}
|p,\Lambda\rangle=\sum_{n,\beta} \int {\rm d}[x]_n\,[{\rm d}^2k_\perp]_n\,\psi^{\Lambda}_{n\,\beta}( r ) |n,x_ip^+,x_i \bm{p}_\perp+\bm{k}_{\perp i}\rangle,\nonumber \\
\end{eqnarray}
where the index $\beta$ refers collectively
to the quark light-front helicities $\lambda_i$, and 
$r$ to the 
momentum coordinates of the quarks in the hadron frame.
In eq.~\eqref{LFWF}, the integration measures are defined as
\begin{equation}
\begin{split}
[\ud x]_n&\equiv\left[\prod_{i=1}^n\ud x_i\right]\delta\!\!\left(1-\sum_{i=1}^nx_i\right),\\
[\ud^2 k_\perp]_n&\equiv\frac{1}{[2(2\pi)^3]^{n-1}}\prod_{i=1}^n \ud^2k_{i\perp}\,\delta^{(2)}\!\!\left(\sum_{i=1}^n k_{i\perp}\right).
\end{split}
\end{equation}
 \\
By inserting~\eqref{operator} and \eqref{LFWF} into the correlator~\eqref{eq:GTMD}, we obtain for the quark contribution
 \ba\label{overlap}
&&W^{[\Gamma]}_{\Lambda'\Lambda}(P,x,\uvec k_\perp,\Delta)=\frac{1}{\sqrt{1-\xi^2}}\sum_n\sum_{\beta',\beta}\int[\ud x]_n\,[\ud^2 k_\perp]_n \nonumber\\
&&\times\bar\delta(\tilde k)\,\psi^{\Lambda'\, *}_{n\beta'}(r')\,\psi^{\Lambda}_{n\beta}(r)\,M^{[\Gamma]\beta'\beta}.
\ea
In the above equation,
 the function $\bar\delta(\tilde k)$ selects the active quark average momentum
\begin{equation*}
\bar\delta(\tilde k)\equiv\sum_{i=1}^n\Theta( x)\,\delta( x-x_i)\,\delta^{(2)}(\uvec k_\perp-\uvec k_{i\perp}).
\end{equation*}
and
\begin{equation}
M^{[\Gamma]\beta'\beta}=\sum_j M^{[\Gamma]\lambda'_j\lambda_j}\,
\prod_{i\neq j}
\delta_{\lambda'_i\lambda_i}, 
\end{equation}
with
 \be
 M^{[\Gamma]\lambda'\lambda}\equiv\frac{\overline{u}(k',\lambda')\Gamma u(k,\lambda)}{2P^+\sqrt{1-\xi^2}}.
 \ee
At leading twist, there are only four  Dirac structures $\Gamma_{\mbox{\footnotesize{tw-2}}} \\=\{\gamma^+,\, i\sigma^{1+}\gamma_5,\, i\sigma^{2+}\gamma_5,\, \gamma^+\gamma_5\}$, which correspond to the four possible transitions of the light-front helicity of the active quark from the initial to the final state
\begin{eqnarray}\label{correspondence}
&&M^{[\gamma^+]\lambda'\lambda}=\delta^{\lambda'\lambda},\nonumber\\
&& M^{[i\sigma^{j+}\gamma_5]\lambda'\lambda}=(\sigma_j)^{\lambda'\lambda},\quad j=1,2\nonumber\\
 && M^{[\gamma^+\gamma_5]\lambda'\lambda}=(\sigma_3)^{\lambda'\lambda}
\end{eqnarray}
with $\sigma_i$ the three Pauli matrices.
\\
Taking into account also all the possible polarization states for the nucleon target, one finds out the quark-quark correlator \eqref{eq:GTMD} can be parametrized in terms of 16 independent 
GTMDs, which are complex valued functions.
Equation \eqref{overlap} provides the model independent LFWF overlap representation of the correlator, which can be easily adopted in different quark models.
Explicit calculations  have been performed in the light-front quark models and in the light-front version of the chiral-quark soliton model ($\chi$QSM)~\cite{Lorce:2011dv}, that will be discussed in more details in sect.~\ref{sec:4} along with other quark model calculations.
\section{Wigner Distributions}
\label{sec:2}

The concept of Wigner distributions in QCD for quarks and gluons was first explored in refs.~\cite{Ji:2003ak,Belitsky:2003nz}. 
Neglecting relativistic effects, the authors used the standard three-dimensional Fourier transform of the GTMDs  in the Breit frame and introduced six-dimensional Wigner distributions (three position and three momentum coordinates). A more natural interpretation in the infinite-momentum frame (IMF)  has been reconsidered in ref.~\cite{Lorce:2011kd}, introducing the definition of
five-dimensional Wigner distributions (two position and three momentum coordinates), which  is not spoiled by relativistic corrections. 
 \\ Following ref.~\cite{Lorce:2011kd}, we define  the quark  Wigner distributions  as 
\begin{eqnarray}\label{wigner}
&&\rho^{[\Gamma]}_{\Lambda',\Lambda}(x,\uvec k_\perp,\uvec b_\perp,{\cal W})\nonumber\\
&&\equiv\int\frac{\ud^2\Delta_\perp}{(2\pi)^2}\,\langle p^+,\tfrac{\uvec\Delta_\perp}{2},\Lambda'|\widehat W^{[\Gamma]}(x, \uvec k_\perp,\uvec b_\perp,{\cal W})|p^+,-\tfrac{\uvec\Delta_\perp}{2},\Lambda\rangle,\nonumber\\
&&
\end{eqnarray}
where the Wigner operator $\widehat W^{[\Gamma]}$ at a fixed light-cone time $z^+=0$ reads 
\begin{eqnarray}\label{wigner-operator}
&&\widehat W^{[\Gamma]}(x,\uvec k_\perp,\uvec b_\perp,{\cal W})\equiv\frac{1}{2}\int\frac{\ud z^-\,\ud^2z_\perp}{(2\pi)^3}\,e^{i(xP^+z^--\uvec k_\perp\cdot\uvec z_\perp)}\,\nonumber\\
&&\times \overline{\psi}(y-\tfrac{z}{2})\Gamma\mathcal W\,\psi(y+\tfrac{z}{2})\big|_{z^+=0}
\end{eqnarray}
with $y^\mu=[0,0,\uvec b_\perp]$.
Note that $\uvec b_\perp$ and $\uvec k_\perp$ are not Fourier conjugate variables, like in the usual quantum-mechanical Wigner distributions. Rather, if $\uvec r_{i\perp}$ ($\uvec r_{f\perp}$) and $\uvec k_{i\perp}$ ($\uvec k_{f\perp}$) are the transverse position and momentum coordinates of the initial (final) quark operator $\psi$ ($\overline\psi$), one sees that the average quark momentum $\uvec k_\perp=(\uvec k_{f\perp}+\uvec k_{i\perp})/2$ is the Fourier conjugate variable of the quark displacement $\uvec z_\perp=\uvec r_{i\perp}-\uvec r_{f\perp}$ and the momentum transfer to the quark $\uvec q_\perp=\uvec k_{f\perp}-\uvec k_{i\perp}$ is the Fourier conjugate variable of the average quark position $\uvec b_\perp=(\uvec r_{f\perp}+\uvec r_{i\perp})/2$. Even though $\uvec b_\perp$ and $\uvec k_\perp$ are not Fourier conjugate variables, they are subjected to Heisenberg's uncertainty principle,
because the corresponding quantum-mechanical operators do not commute $[\hat{\uvec b}_\perp,\hat{\uvec k}_\perp]\neq 0$. 
\begin{figure*}[th!]
	\centering
		\includegraphics[width=.49\textwidth]{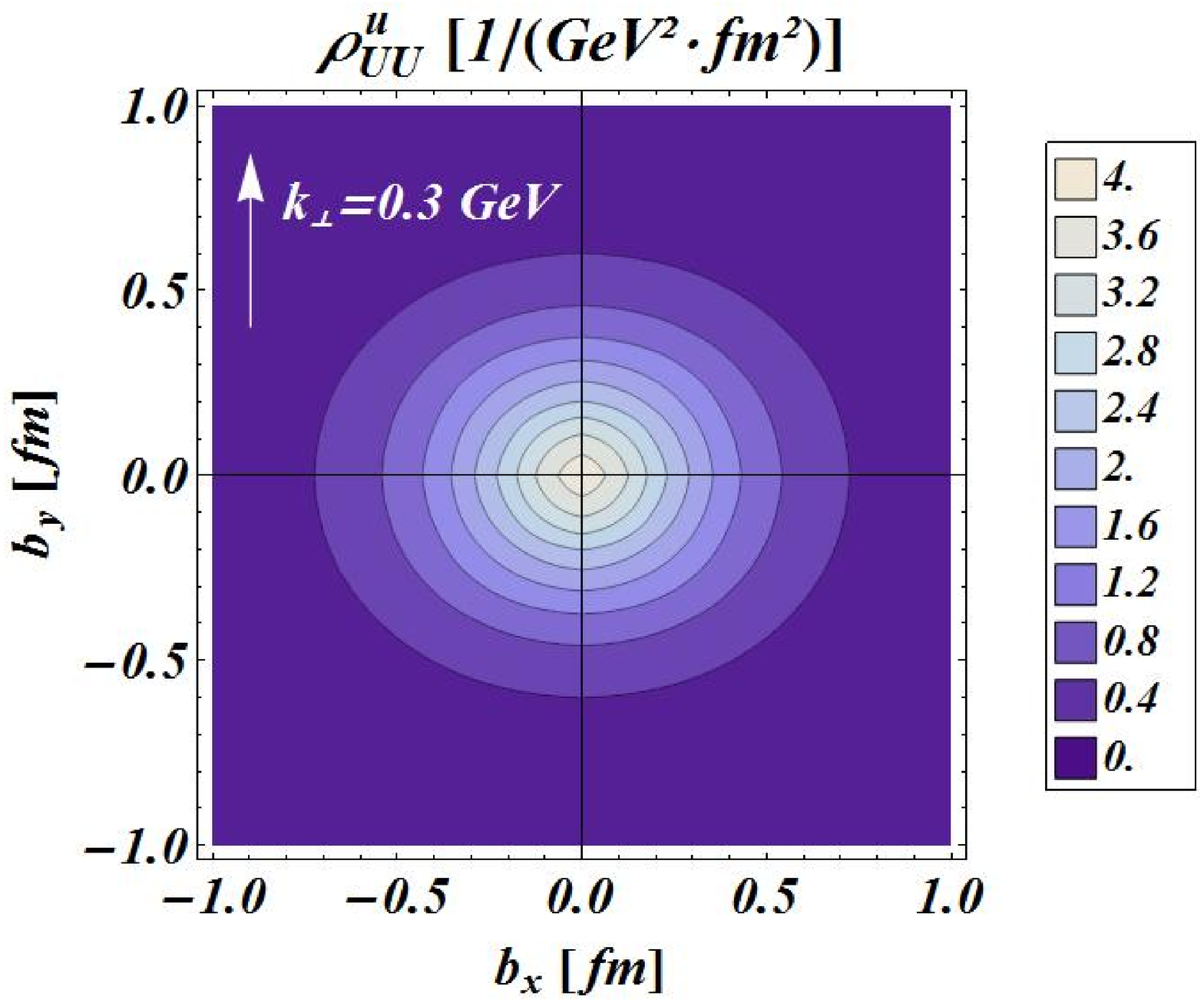}
		\includegraphics[width=.49\textwidth]{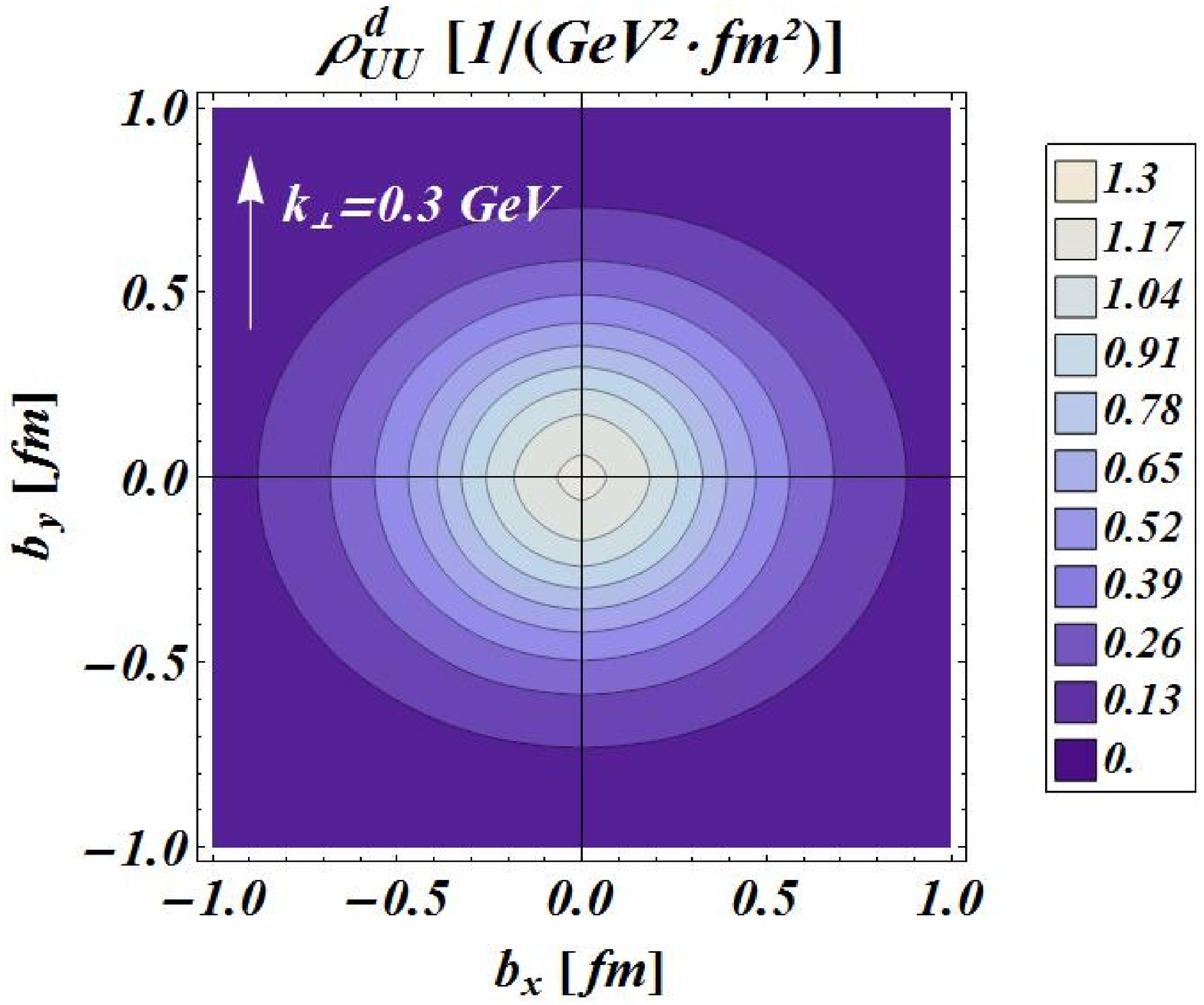}
\caption{\footnotesize{(Color online) The $x$-integrated Wigner distributions of unpolarized  quarks in an unpolarized proton in impact-parameter space with fixed transverse momentum $\uvec k_\perp=k_\perp\,\hat e_y$ and $k_\perp=0.3$ GeV. The left (right) panel shows the results for $u$ ($d$) quarks.
% (Picture from ref.~\cite{Lorce:2011kd}).
}}
		\label{fig2}
\end{figure*}
\\
Thanks to the properties of the Galilean subgroup of transverse boosts in the IMF \cite{Burkardt:2005td,Kogut:1969xa}, we can form a localized nucleon state in the transverse direction, in the sense that its transverse center of momentum is at the position $\uvec r_\perp$ :
\begin{equation}
|p^+,\uvec r_\perp\rangle=\int\frac{\ud^2p_\perp}{(2\pi)^2}\,e^{-i\uvec p_\perp\cdot\uvec r_\perp}|p^+,\uvec p_\perp\rangle.
\end{equation}
The Wigner distributions in eq.~\eqref{wigner} can then be written in terms of these localized nucleon states as
\begin{eqnarray}\label{wignerloc}
&&\rho^{[\Gamma]}_{\Lambda',\Lambda}(x, \uvec k_\perp, \uvec b_\perp,{\cal W})\nonumber\\
&&=\int\ud^2D_\perp\,\langle p^+,-\tfrac{\uvec D_\perp}{2},\Lambda'|\widehat W^{[\Gamma]}(x,\uvec k_\perp,\uvec b_\perp,{\cal W})|p^+,\tfrac{\uvec D_\perp}{2},\Lambda\rangle,\nonumber\\
&&
\end{eqnarray}
where $\uvec D_\perp$ is the transverse displacement between the initial and final centers of momentum. From eqs.~\eqref{wigner} and \eqref{wignerloc}, we also note
 that the nucleon state has vanishing average transverse position and average transverse momentum. This allows us to interpret the variables $\uvec b_\perp$ and $\uvec k_\perp$ as the relative average transverse position and the relative average transverse momentum of the quark, respectively. 

Using a transverse translation in eq.~\eqref{wigner}, we obtain the Wigner distributions as  two-dimensional Fourier transforms of the GTMDs  correlator~\eqref{eq:GTMD} for $\Delta^+=0$, {\it i.e.}
\begin{eqnarray}
&&\rho^{[\Gamma]}_{\Lambda',\Lambda}(x,\uvec k_\perp,\uvec b_\perp,{\cal W})\nonumber\\
&&=\int\frac{\ud^2\Delta_\perp}{(2\pi)^2}\,e^{-i\uvec\Delta_\perp\cdot\uvec b_\perp} \nonumber\\
&&\times\langle p^+,-\tfrac{\uvec \Delta_\perp}{2},\Lambda'|\widehat W^{[\Gamma]}(x,\uvec k_\perp,\uvec 0_\perp,{\cal W})|p^+,\tfrac{\uvec \Delta_\perp}{2},\Lambda\rangle\nonumber\\
&&=
\int\frac{\ud^2\Delta_\perp}{(2\pi)^2}\,e^{-i\uvec\Delta_\perp\cdot\uvec b_\perp} W^{[\Gamma]}_{\Lambda'\Lambda}(P,x,\uvec k_\perp,\uvec\Delta_\perp,{\cal W}),
\end{eqnarray}
 Note that the Hermiticity property of the GTMD correlator 
 \begin{equation}
  [W^{[\Gamma]}_{\Lambda'\Lambda}(P,x,\uvec k_\perp,\uvec\Delta_\perp,{\cal W})]^*= W^{[\Gamma]}_{\Lambda'\Lambda}(P,x,\uvec k_\perp,-\uvec\Delta_\perp,{\cal W})
  \end{equation}
  implies that the Wigner distributions are real quantities, in accordance with their
interpretation as a phase-space distributions. However, because of the uncertainty principle which prevents knowing simultaneously the position and momentum of a quantum-mechanical system, they do not have a strict probabilistic interpretation. To overcome this problem, recently it has been proposed to use  Husimi distributions~\cite{Hagiwara:2014iya}, which have the advantage of being positive-definite, but do not have  a direct link to measurable quantities such as GPDs and TMDs.
 
 Predictions  for the quark Wigner distributions have been obtained 
in phenomenological and perturbative  models, such as a light-front constituent quark model (LFCQM) \cite{Lorce:2011kd}, quark-target model~\cite{Kanazawa:2014nha,Mukherjee:2014nya}, and a light-front quark-diquark model \cite{Liu:2014vwa}. In all these models the role of the Wilson line has been neglected, and the calculation has been performed taking the Fourier transform of the GTMDs obtained  in terms of overlap of LFWFs as outlined in sect.~\ref{sec:1}. As an example, we will discuss the results in a LFCQM, which allows one to
sketch some general features about the behavior
of the quarks in the nucleon when observed in the phase-space. 
In fig. \ref{fig2} we show  the $x$-integrated Wigner distributions $\rho_{UU}=\tfrac{1}{2} \sum_{\Lambda}\rho^{[\gamma^+]}_{\Lambda,\Lambda}$ for unpolarized quarks
in an unpolarized nucleon in impact-parameter space with fixed transverse momentum $\uvec k_\perp=k_\perp \hat e_y$ with $k_\perp=0.3$ GeV.
The left (right) panel corresponds to the $u$ ($d$) quark contribution. 
These distributions show a nontrivial correlation between the transverse position and momentum of the quarks.
They are 
are not axially symmetric, showing  that the configuration
$\uvec b_\perp \perp \uvec k_\perp$ is favored with respect to the configuration $\uvec b_\perp \parallel \uvec k_\perp$.
This deformation can be explained as a consequence of confinement which limits
the radial motion of the quark with respect to its orbital
motion. Furthermore, 
the distribution for $u$ quarks is more concentrated at the center of the proton, while the $d$-quark distribution extends further at the periphery of the proton.
The left-right symmetry of the distributions is a consequence of a vanishing net orbital angular momentum, indicating that the quarks have no preference on moving either
clockwise or counterclockwise. The top-bottom symmetry instead is not a general property of the $\rho_{UU}$ distribution, but it is due to the fact that we ignored  the contribution of the Wilson line representing the effect of initial and final state interactions. 
\\%
\begin{figure*}[t!]
\resizebox{\textwidth}{!}{
\includegraphics[width=0.5\textwidth]{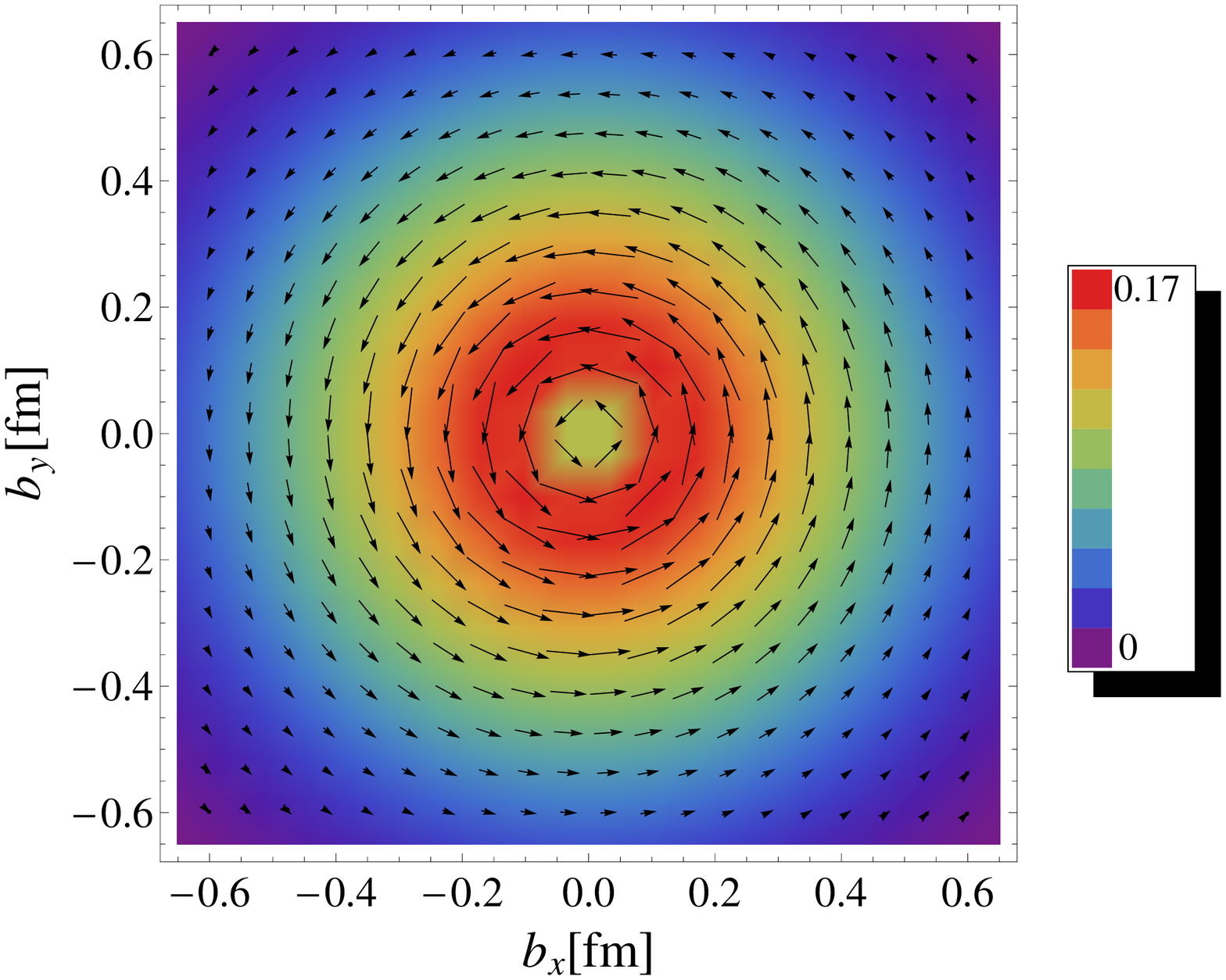}
\includegraphics[width=0.5\textwidth]{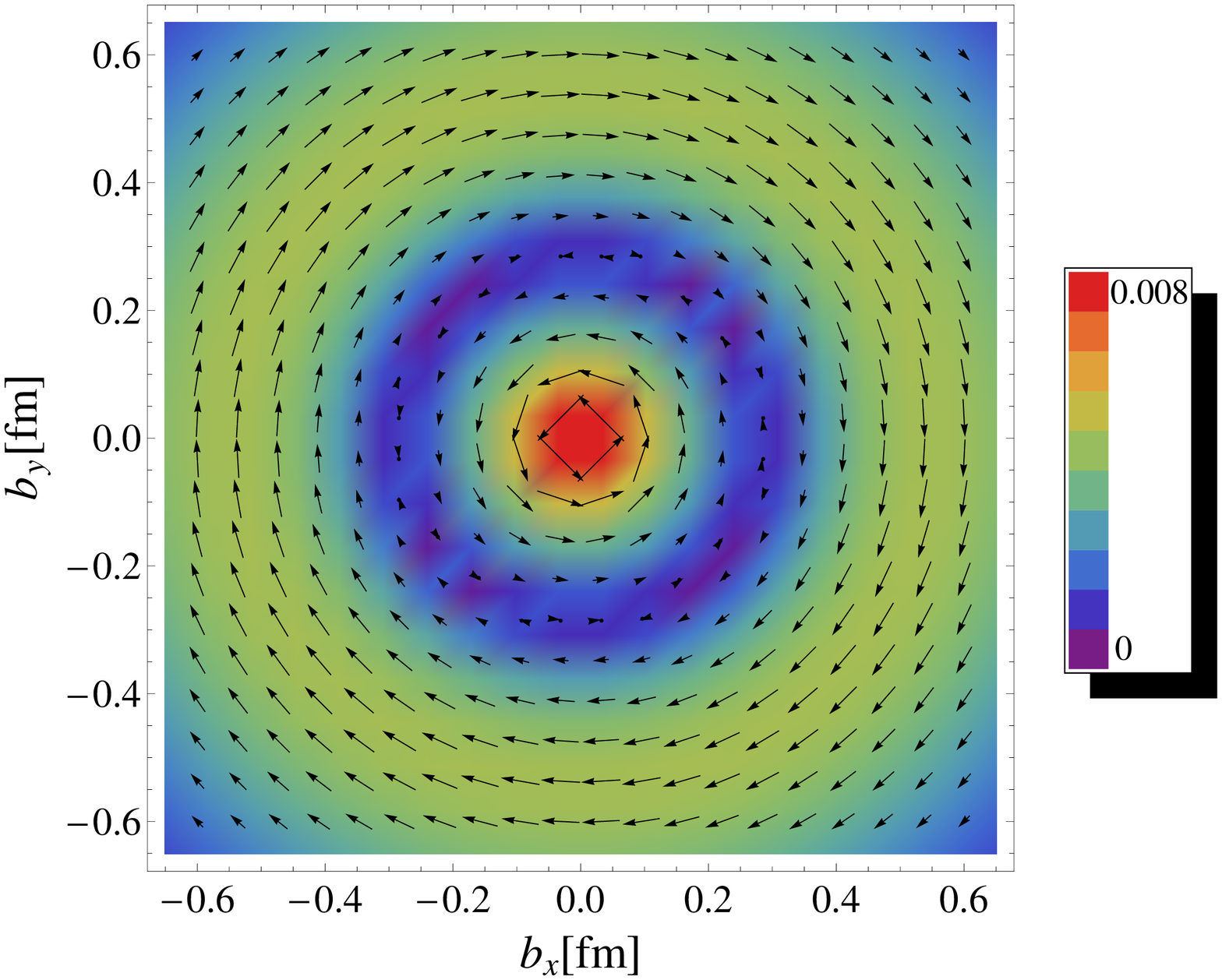}
}
\caption{(Color online) Distributions in impact-parameter space of the mean transverse momentum of unpolarized quarks in a longitudinally polarized nucleon. The nucleon polarization is pointing out of the plane, while the arrows show the size and direction of the mean transverse momentum of the quarks. The left (right) panel shows the results for $u$ ($d$) quarks. 
%(Picture from ref.~\cite{Lorce:2011ni}.)
}
\label{fig3} \vspace{-0mm}
\end{figure*}
In the case of polarized  quarks and/or gluons, one can also observe nontrivial spin-orbit correlations  in the phase-space~\cite{Lorce:2011kd,Lorce:2014mxa}.
Here we will focus on the Wigner distribution for unpolarized quarks in a longitudinally polarized nucleon, $\rho^{[\gamma^+]}_{+,+}$,
which  provides an intuitive and simple representation of the quark
OAM~\cite{Lorce:2011kd,Lorce:2011ni}.
Such a relation resembles the classical formula given by the
phase-space average of the OAM
weighted by the density operator, and reads
\begin{eqnarray}
\label{oam:wigner}
{\cal L}_z=\int\ud x\,\ud^2k_\perp\,\ud^2\uvec b_\perp (\uvec b_\perp\times \uvec k_\perp)_z\,\rho^{[\gamma^+]}_{+,+}(x,\uvec k_\perp,\uvec b_\perp,{\cal W}).\nonumber\\
\end{eqnarray}
Introducing the distribution of average quark transverse momentum in phase-space  
\begin{eqnarray}
\langle \uvec k_\perp\rangle (\uvec b_\perp)=\int\ud x\,\ud^2k_\perp\, \uvec k_\perp\,\rho^{[\gamma^+]}_{+,+}(x,\uvec k_\perp,\uvec b_\perp,{\cal W}),\label{k:wigner}
\end{eqnarray}
we can also rewrite eq.~\eqref{oam:wigner} as
\begin{eqnarray}
\label{oam:wigner2}
{\cal L}_z=\int \ud^2 b_\perp [\uvec b_\perp\times\langle {\uvec k}_\perp\rangle(\uvec b_\perp)]_z.
\end{eqnarray}
In fig.~\ref{fig3} we show
the results for the distribution of average quark transverse momentum from a LFCQM. Since we are neglecting the contribution of the Wilson line, we observe that
 the mean transverse-momentum $\langle \uvec k_\perp\rangle$
is always orthogonal to the impact-parameter vector $\uvec b_\perp$.
The orbital motion of the $u$ quark corresponds to an OAM in the positive $\hat z$ direction, with decreasing magnitude towards the periphery.
For $d$ quarks we observe two regions. In the
central region of the nucleon the $d$ quarks
tend to orbit counterclockwise like the $u$ quarks, while in the
peripheral region they  tend to orbit
clockwise, leading to a very small net OAM.
 Although this behavior is specific to the assumed model, fig.~\ref{fig3} provides an instructive illustration of the kind of information one can learn about the density of the OAM in impact-parameter space. 

\section{Transverse Force/Torque on Quarks in DIS}
\label{sec:3}
So far we did not discuss the role of the Wilson-line gauge link in the Wigner functions. This raises the question about the path to be chosen for the gauge link, as the choice of path matters when there are nonzero 
color-electric and -magnetic fields.
When one evaluates averages of ${\uvec k}_\perp$, such as eqs. \eqref{k:wigner} and \eqref{oam:wigner2}, the only other term involving  ${\uvec k}_\perp$ is the factor $e^{-{\uvec k}_\perp\cdot{\uvec z}_\perp}$
in eq. (\ref{wigner-operator}), i.e. one can replace ${\uvec k}_\perp$  first by a gradient 
$i{\uvec \nabla}_{{\uvec z}_\perp}$ and after integration by parts by a gradient 
$-i{\uvec \nabla}_{{\uvec z}_\perp}$ acting on $\bar{\psi}(y-\frac{z}{2})\Gamma {\cal W}\psi(y+\frac{z}{2})$.
This results in the canonical momentum operator acting on the quark fields plus the derivative acting on 
${\cal W}$. Finally due to $\int d^2k_\perp e^{-i{\uvec k}_\perp\cdot{\uvec z}_\perp}=(2\pi)^2\delta^{(2)}({\uvec z}_\perp)$
the operator becomes local in ${\uvec z}_\perp$.

One natural choice for the path is a straight line resulting in\footnote{We stick to the notation of sect.~\ref{sec:2},  where $y^\mu=[0,0,\vec b_\perp]$.}
\begin{eqnarray}
&&-i{\bf \nabla}_{{\uvec z}_\perp}\left.\bar{\psi}(y-\frac{z}{2})\gamma^+ {\cal W}_{straight}\psi(y+\frac{z}{2})
\right|_{{\uvec z}_\perp=0}
\nonumber\\
&&\quad =\quad 
\bar{\psi}(y)\gamma^+ \left[-i\stackrel{\leftrightarrow}{{\bf \nabla}}_{{\uvec y}_\perp}
+g{\uvec A}_\perp (y)\right]\!
\psi(y) . \label{eq:kJi}
\end{eqnarray}
For the OAM this results in
\begin{equation}
-i\bar\psi(y)\gamma^+\left[\uvec y\times\stackrel{\leftrightarrow}{\uvec D}_{{\uvec y}_\perp}\right]_z\psi(y),
\label{ji:oam}
\end{equation}
where ${\uvec D}_{{\uvec y}_\perp}\equiv {\bf \nabla}_{{\uvec y}_\perp}+ig{\uvec A}_\perp(y)$,
which is identical to the quark OAM in the Ji decomposition~\cite{Ji:1996ek}. However, for the average transverse momentum one obtains
a vanishing result since the operator in eq. \eqref{eq:kJi} is odd under PT.
Another, more intuitive argument is that the intrinsic average transverse momentum of quarks relative to that of a nucleon must vanish since the quarks are bound and not flying away.

\begin{figure}[t]
\resizebox{0.5\textwidth}{!}{
\includegraphics[width=0.5\textwidth]{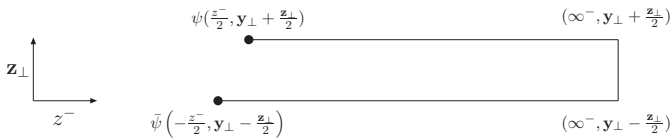}
}
\caption{Illustration of the path for the Wilson line gauge link connecting 
$\bar{\psi}(y-\frac{z}{2})$ and $q(y+\frac{z}{2})$
in the definition of the Wigner function that includes FSIs.}
\label{fig:staple}
\end{figure}

Alternatively, one can include final state interactions (FSIs) in eikonal approximation into the definition of Wigner functions by including gauge links along the $z^-$ direction from the position of the quark field
operator to $\infty$ - the trajectory of an ejected quark in a DIS experiment. Physically this is equivalent
to measuring physical observables such as $\langle {\uvec k}_\perp \rangle({\uvec b}_\perp)$ or 
${\cal L}_z$ not at the original position inside the hadron but after it has been 
ejected in a DIS experiment. In order to ensure gauge invariance one has to close the connection between $\psi$ and
$\bar{\psi}$ at $\infty$ but the exact shape of that segment does not matter since the field strength tensor vanishes at $\infty$ (fig. \ref{fig:staple}).

In light-cone gauge $A^+=0$ only the segment at $z^-=\infty$ contributes and one finds
for the operator that represents $\langle {\uvec k}_\perp \rangle$
\begin{eqnarray}
& &-i{\bf \nabla}_{{\uvec z}_\perp}\left.
\bar{\psi}(y-\frac{z}{2})\gamma^+ {\cal W}_{staple}\psi(y+\frac{z}{2})\right|_{{\uvec z}_\perp={\bf 0}}
\nonumber\\
& &\quad {\longrightarrow}\quad
\bar{\psi}(y)\gamma^+ \left[-i\stackrel{\leftrightarrow}{{\bf \nabla}}_{{\uvec y}_\perp}+g{\uvec A}_\perp (\infty,{\uvec y}_\perp)\right]
\psi(y) . \label{eq:kJM}\end{eqnarray}
In gauges other than $A^+=0$ there is an additional contribution from the light-like Wilson lines 
and one finds\footnote{For simplicity, only the result for abelian fields is shown here. In the nonabelian
case additional Wilson lines appear \cite{Burkardt:2008ps}.}
\begin{eqnarray}\hspace{-0.8 truecm}
{\uvec A}_\perp (\infty,{\uvec y}_\perp) &\rightarrow& {\uvec A}_\perp (\infty,{\uvec y}_\perp) 
- {\bf \nabla}_\perp \int_{y^-}^\infty\!dr^- A^+(r^-,{\uvec y}_\perp)\nonumber\\
\hspace{-0.8 truecm}&=&{\uvec A}_\perp (y^-,{\uvec y}_\perp) +\int_{y^-}^\infty \!dr^-{{\uvec G}^{+}}_\perp(r^-,{\uvec y}_\perp),
\label{eq:force}
\end{eqnarray}
where ${{\uvec G}^{+}}_\perp$ are components of the gluon field strength tensor, i.e.
\begin{eqnarray}
\sqrt{2}g{G^+}_x &=& g\left(E_x+B_y\right),\nonumber\\
\sqrt{2}g{G^+}_y &=& g\left(E_y-B_x\right).\label{eq:Lorentz}
\end{eqnarray}
The operator representing $\langle {\uvec k}_\perp ({\uvec b}_\perp)\rangle$ with FSI included
then reads 
\begin{equation}
\bar{\psi}(y)\gamma^+\left[-i\stackrel{\leftrightarrow}{\uvec D}_{{\uvec y}_\perp}+\int_{y^-}^\infty \!\!dr^-g{{\uvec G}^{+}}_\perp(r^-,{\uvec y}_\perp)
\right]\psi(y).\label{eq:kofb}
\end{equation}
As observed above, the expectation value of
$
\bar{\psi}(y)\gamma^+\stackrel{\leftrightarrow}{\uvec D}_{{\uvec y}_\perp}\psi(y)
$
vanishes when integrated over ${\uvec y}_\perp={\uvec b}_\perp$ since it is PT odd. Therefore, only the second term contributes to the average transverse momentum of the quarks, which is thus given by the expectation value of the operator
\begin{equation}
\bar{\psi}(y)\gamma^+\left[-g\int_{y^-}^\infty \!\!dr^-{\uvec G}^{+\perp}(r^-,{\uvec y}_\perp)
\right]\psi(y)
\label{eq:trans-staple}
\end{equation}
integrated over the transverse position. This result has a very intuitive interpretation.
For a particle moving with
the speed of light in the $-\hat{z}$ direction $r^3=-r^0$ and thus $dr^-=\sqrt{2}dr^0$. 
Furthermore, from eq. \eqref{eq:Lorentz}, we note that 
\begin{equation}
-\sqrt{2}g{\uvec G}^{+\perp}=g\left[{\uvec E}_\perp+({\vec v}\times{\vec B})_\perp\right]
\end{equation}
represents the color Lorentz force acting on a particle that moves with the velocity of light  in the $-\hat{z}$
direction. Therefore, $-g\int_{y^-}^\infty \!\!dr^-{\uvec G}^{+\perp}(r^-,{\uvec y}_\perp)$ represents the color
Lorentz force integrated over time.
This observation motivates the
semi-classical interpretation of the matrix element \eqref{eq:trans-staple} as the average transverse momentum of the ejected quark in a DIS experiment
as arising from the average color-Lorentz force from the spectators as it leaves the target, which is an example for Ehrenfest's theorem applied to QCD.
This should not be entirely surprising since the light-like Wilson line was introduced to include the FSIs.

Another application of eq. \eqref{eq:kofb} is to the quark orbital angular momentum \eqref{oam:wigner2}. 
Using light-cone staples for the gauge links in light-cone gauge gives rise to the canonical expression plus a term at $z^-=\infty$. Such a term was not included in \cite{Jaffe:1989jz}. In ref. \cite{Bashinsky:1998if}, 
a 'zero mode' term was included that fixed any issues with residual gauge invariance and contributions from
${\uvec A}_\perp(z^-=\pm \infty, {\uvec y}_\perp)$.
It turns out that for antisymmetric boundary conditions the term at $z^-=\infty$ does not matter for the definition of OAM and thus in this case  the definition of quark OAM using light-cone staples is equivalent to the Jaffe-Manohar-definition \cite{Jaffe:1989jz}. For simplicity, we will here assume that such boundary conditions have been imposed in
which case the OAM with light-cone staple gauge links is equivalent to the Jaffe-Manohar definition.

With this in mind, the difference between the operators entering the Ji vs. Jaffe-Manohar OAM reads
\begin{eqnarray}
\bar\psi(y)\gamma^+ \int_{y^{-}}^\infty \frac{{\rm d}r^-}{\sqrt{2}} T^z (r^-,\uvec y_\perp)\psi(y),
\end{eqnarray}
where 
\begin{equation}
T^z (r^-,\uvec r_\perp)=-\sqrt{2}g\left(xG^{+y}(r^-,\uvec r_\perp)-yG^{+x}(r^-,\uvec r_\perp)\right)
\end{equation}
represents the $\hat z$ component of the torque that acts on a particle moving with (nearly) the velocity of light  in the
$-\hat z$ direction \cite{Burkardt:2012sd}.
Therefore, Ji's OAM represents the local and manifestly gauge invariant OAM of the
quark {\it before} it has been struck by the virtual photon in a DIS experiment, while the Jaffe-Manohar OAM is
the gauge invariant OAM {\it after} it has left the nucleon and moved to $r^-=\infty$. In other words,
their difference represents the change in OAM as the quark leaves the target.
While it had been demonstrated earlier that the difference between Jaffe-Manohar's and Ji's OAM is nonzero
\cite{Burkardt:2008ua}, the physical interpretation of their difference was only explained later
\cite{Burkardt:2012sd}.

\begin{figure}[t]
\resizebox{0.5\textwidth}{!}{
\includegraphics[width=0.5\textwidth]{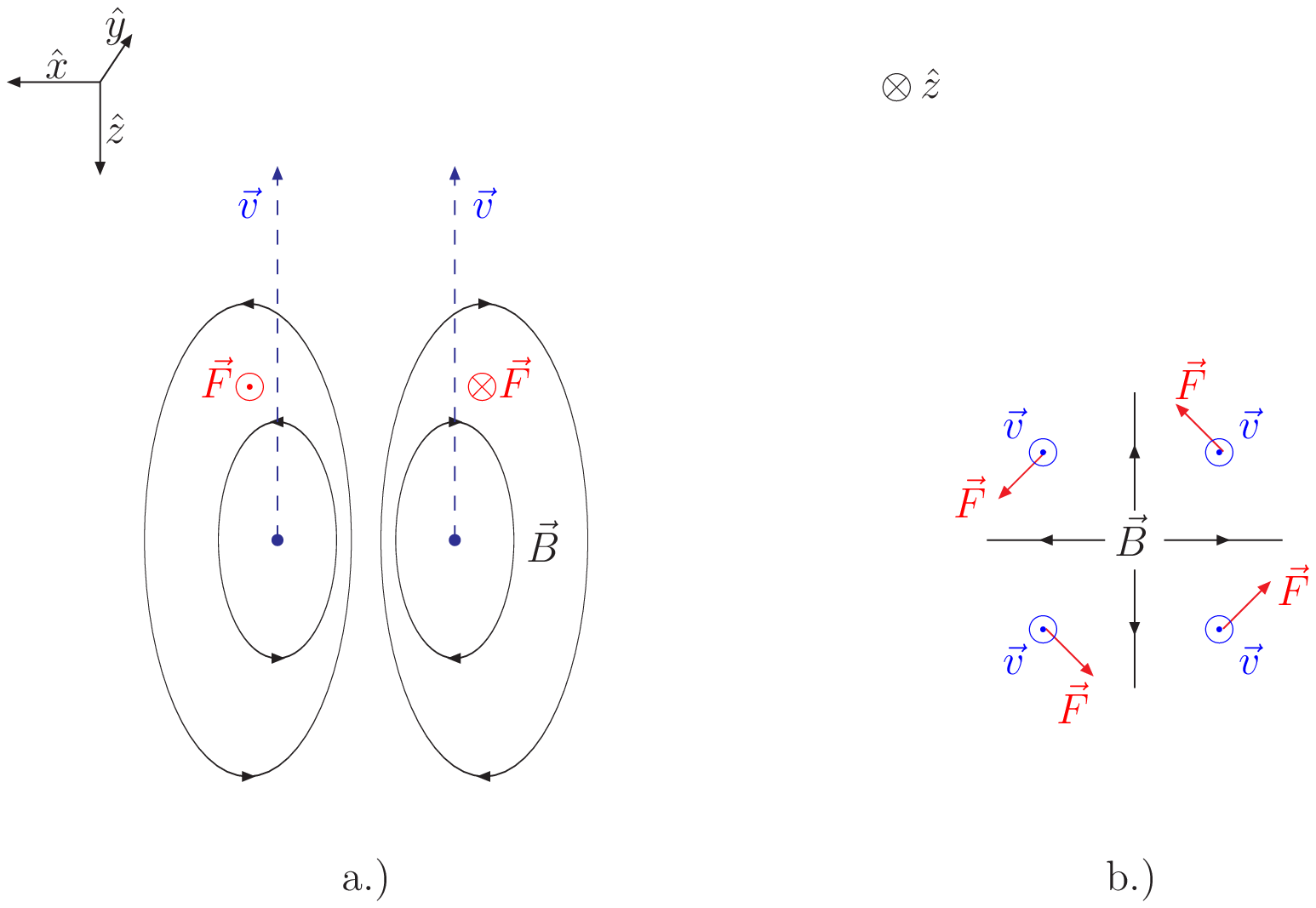}
}\caption{(Color online) Illustration of the torque acting on a positron moving in the $-\hat{z}$ direction
through a magnetic dipole field caused by the magnetic moment of an electron polarized in the
$+\hat{z}$ direction. a.) side view; b.) top view. 
In this example the $\hat{z}$ component of the torque is negative as the positron leaves the
bound state.
}
\label{fig:dipole}
\end{figure}

In order to estimate the effect from the final state interactions on the quark OAM we first consider the effect
on a positron moving through the magnetic dipole field of an electron, which is polarized in the $+\hat{z}$ direction. This should be the most simple analog to a proton polarized in the $+\hat{z}$ direction 
because more quarks are polarized in the same direction as the nucleon spin and the color-electric force between the active quark and the spectators is attractive. As illustrated in fig. \ref{fig:dipole}, the torque along the trajectory should on average be negative. This would suggest that the Jaffe-Manohar OAM is less than the Ji OAM.
Preliminary lattice QCD calculations are consistent with this simple picture \cite{Engelhardt}.

\section{Transverse-momentum dependent Parton Distributions}
\label{sec:4}

The forward limit $\Delta=0$ of the correlator \eqref{eq:GTMD} is given by the following quark-quark correlator, denoted as $\Phi$~\cite{Kotzinian:1994dv,Mulders:1995dh}
\begin{eqnarray}
&&\Phi^{[\Gamma]}_{\Lambda'\Lambda}(P,x,\uvec k_\perp,{\cal W})=W^{[\Gamma]}_{\Lambda'\Lambda}(P,x,\uvec k_\perp,0,{\cal W})\qquad\qquad\qquad\nonumber\\
&&=\int\frac{\ud z^-\,\ud^2z_\perp}{2(2\pi)^3}\,e^{ixP^+z^--i\uvec k_\perp\cdot\uvec z_\perp}\nonumber\\
&&\times\langle p,\Lambda'|\overline\psi(-\tfrac{z}{2})\Gamma\,\mathcal W\,\psi(\tfrac{z}{2})|p,\Lambda\rangle\Big|_{z^+=0}.
\label{corrTMD}
\end{eqnarray}

It is convenient to represent the correlator \eqref{corrTMD} in the four-component basis by the tensor $\Phi^{\mu\nu}$~\cite{Lorce:2011zta}. This tensor is related to helicity amplitudes as follows
\begin{eqnarray}\label{basischange}
&&\Phi^{\mu\nu}=\frac{1}{2}\sum_{\Lambda'\Lambda\lambda'\lambda}(\bar\sigma^\mu)^{\Lambda\Lambda'}(\bar\sigma^\nu)^{\lambda'\lambda}\,\Phi_{\Lambda'\lambda',\Lambda\lambda},\nonumber\\
&&\Phi_{\Lambda'\lambda',\Lambda\lambda}=\frac{1}{2}\,\Phi^{\mu\nu}(\sigma_\mu)_{\Lambda'\Lambda}(\sigma_\nu)_{\lambda\lambda'},
\end{eqnarray}
where $\bar\sigma^\nu=(\mathds{1},\usigma)$ and  $\sigma^\nu=(\mathds{1},-\usigma)$. 
The tensor correlator is then given by the following combinations of TMDs
\begin{align}
\Phi^{\mu\nu}&=\begin{pmatrix}
f_1&\frac{k_y}{M}\,h^{\perp }_1&-\frac{k_x}{M}\,h^{\perp }_1&0\\
\frac{k_y}{M}\,f^{\perp }_{1T}&h_1+\frac{k^2_x-k^2_y}{2M^2}\,h^{\perp }_{1T}&\frac{k_xk_y}{M^2}\,h^{\perp }_{1T}&\frac{k_x}{M}\,g_{1T}\\
-\frac{k_x}{M}\,f^{\perp }_{1T}&\frac{k_xk_y}{M^2}\,h^{\perp }_{1T}&h_1-\frac{k^2_x-k^2_y}{2M^2}\,h^{\perp }_{1T}&\frac{k_y}{M}\,g_{1T}\\
0&\frac{k_x}{M}\,h^{\perp }_{1L}&\frac{k_y}{M}\,h^{\perp }_{1L}&g_{1L}
\end{pmatrix},\label{TMDs}
\end{align}
where we introduced the notations $h_{1T}^{\pm }=h_1\pm\tfrac{\uk_\perp^2}{2M^2}\,h_{1T}^{\perp }$ and $\hat k_i=k_i/k_\perp$ with $k_\perp=|\uk_\perp|$. 
Out of the eight TMDs in \eqref{TMDs},  the so-called Boer-Mulders function $h_{1}^{\perp }$  \cite{Boer:1997nt}
 and the Sivers function $f_{1T}^{\perp }$ \cite{Sivers:1989cc} are T-odd,
 i.e. they change sign under naive time reversal, which is defined as usual
 time-reversal but without interchange of initial and final states, while the other six are T-even.
The different components of the tensor correlator \eqref{TMDs} correspond to quark momentum density for different light-front polarization states  of the quarks and the target.
The component $\Phi^{00}$  gives the density of unpolarized quarks in the unpolarized target. The components $\Phi^{0j}$ give the net density of quarks with light-front polarization in the direction $\boldsymbol e_j$ in the unpolarized target, while the components $\Phi^{i0}$ give the net density of unpolarized quarks in the target with light-front polarization in the direction $\boldsymbol e_i$. Finally, the components $\Phi^{ij}$ give the net density of quarks with light-front polarization in the direction $\boldsymbol e_j$ in the target with light-front polarization in the direction $\boldsymbol e_i$. The density of quarks with definite light-front polarization in the direction $\us$ inside the target with definite light-front polarization in the direction $\uS$ is then  given by $\Phi(x,\uk_\perp,\uS,\us)=\frac{1}{2}\,\bar S_\mu\Phi^{\mu\nu}\bar s_\nu$, where we have introduced the four-component vectors $\bar S_\mu=(1,\uS)$ and $\bar s_\nu=(1,\us)$.\\
TMDs are functions of $x$ and $\uvec k_\perp^2$. The multipole pattern in $\uvec k_\perp$ is clearly visible in eq. \eqref{TMDs}. The TMDs $f_1$, $g_{1L}$ and $h_1$ give the strength of monopole contributions and correspond to matrix elements without a net change of helicity between the initial and final states. The TMDs $f^\perp_{1T}$, $g_{1T}$, $h^\perp_1$, and $h^\perp_{1L}$ give the strength of dipole contributions and correspond to matrix elements involving one unit of helicity flip, either on the nucleon side ($f^\perp_{1T}$ and $g_{1T}$) or on the quark side ($h^\perp_1$ and $h^\perp_{1L}$). Finally, the TMD $h^\perp_{1T}$ gives the strength of the quadrupole contribution and corresponds to matrix elements where both the nucleon and quark helicities flip, but in opposite directions.
Conservation of total angular momentum tells us that helicity flip is compensated by a change of orbital angular momentum \cite{Pasquini:2008ax} which manifests itself by powers of $\uvec k_\perp/M$. 
\newline
TMDs were studied in various low-energy QCD-inspired models. From the point of view of modelling, one should distinguish between T-even and T-odd TMDs.
The former can be studied in models with only quark degrees of freedom, while T-odd TMDs require explicit gauge degrees of freedom.
\\
Models for T-even TMDs can be grouped in the following categories:  (a) light-front quark models~\cite{Pasquini:2008ax,Boffi:2009sh,She:2009jq,Zhu:2011zza,Muller:2014tqa}; (b) spectator models \cite{Jakob:1997wg,Goldstein:2002vv,Gamberg:2003ey,Gamberg:2007wm,Bacchetta:2008af,Lu:2010dt}; (c) chiral quark soliton models~\cite{Lorce:2011dv,Wakamatsu:2009fn,Schweitzer:2012hh}; (d) bag models~\cite{Avakian:2008dz,Avakian:2010br}; (e) covariant parton models~\cite{Efremov:2009ze,Efremov:2010mt}; (f) quark-target model \cite{Meissner:2007rx,Kundu:2001pk,Goeke:2006ef,Mukherjee:2009uy}.
All  these models, with the exception of the covariant parton model and the applications with phenomenological LFWFs of ref.~\cite{Muller:2014tqa}, provide predictions at some low scale, below 1 GeV, that is effectively assumed
to be the threshold between the nonperturbative and perturbative regimes. In order to compare with experimental data, the predictions should be evolved to a higher
scale. Only approximate evolution schemes have been employed for phenomenological studies so far \cite{Boffi:2009sh}, while the application of the correct TMD evolution equations~\cite{Collins:1984kg,Aybat:2011zv} has not been carried out yet.
\\
(a) Light-front models are built by expanding the nucleon state in the basis of free parton (Fock) states. The expansion is truncated to the minimal Fock-state, corresponding to three valence quarks in LFCQM~\cite{Pasquini:2008ax,Boffi:2009sh} or a valence quark and spectator diquark in the light-front quark-diquark model \cite{She:2009jq,Zhu:2011zza,Muller:2014tqa}.   The probability amplitude to find these parton configurations in the nucleon is given by LFWF (see eq.~\ref{LFWF}). 
The LFWF overlap representation of the TMD correlator can be easily obtained from the forward limit of eq.~\eqref{overlap}, {\it i.e.}
\begin{eqnarray}\label{phi-overlap3}
  \Phi^{[\Gamma]}_{\Lambda'\Lambda}(P,x,\uvec k_\perp)&=&
   \sum_{\lambda,\lambda'}
   \frac{ \overline u_{\lambda'}(\tilde{k})\Gamma u_\lambda(\tilde{k})}{2k^+}
   \,{\cal P}^q_{\Lambda'\lambda',\Lambda\lambda}(\tilde k),
   \end{eqnarray}
where 
\begin{eqnarray}
   {\cal P}^q_{\Lambda'\lambda',\Lambda\lambda}(\tilde k)=\sum_n\sum_{\beta,\beta'}\int[\ud x]_n\,[\ud^2k_\perp]_n \bar\delta(\tilde k)\,\psi^{\Lambda'\, *}_{n\beta'} ( r )\,\psi^{\Lambda}_{n\beta}( r ),\nonumber\\
   \label{mom-density1}
   \end{eqnarray}
   with $r_i=(x_i,\uvec k_{\perp  i})$
   and the sum over the $n$-parton states being truncated to $n=3$ in the LFCQM and to $n=2$ in the quark-diquark model.
${\cal P}^q_{\Lambda'\lambda',\Lambda\lambda}$ corresponds to the diagonal components of the quark-momentum density operator, evaluated in the target for different
 quark light-front helicity states, making evident the interpretation of the TMDs as probability densities in the quark-momentum space.
Furthermore, the LFWFs are eigenstates of the total OAM in the light-cone gauge, and therefore allow one to map in a transparent way the multipole pattern in $\uvec k_\perp$ associated with each TMD.
\\
In a recent work~\cite{Lorce:2014hxa}, the application of the LFCQM has been  extended also to higher-twist TMDs.
  
(b) The basic idea of spectator models is to evaluate the
TMD correlator \eqref{corrTMD} by inserting a complete
set of intermediate states and then truncating this set at tree
level to a single on-shell spectator diquark state, i.e. a state
with the quantum numbers of two quarks. The spectator can have the quantum numbers of a scalar (spin 0) isoscalar or axial-vector (spin 1) isovector diquark, and represents an effective particle which combines non-perturbative effects related to the sea and gluon content of the nucleon. Spectator models differ by their specific choice of target-quark-diquark vertices, polarization four-vectors associated with the axial-vector diquark, and vertex form factors. 
The calculation of TMDs within the quark-diquark spectator model can  be  conveniently recast also in the language of two-body LFWFs~\cite{Brodsky:2000ii}, using the formalism outlined above.
A drawback of the model  is that momentum and quark-number sum rules
cannot be satisfied simultaneously, unless one considers the diquark as composed of two
quarks~\cite{Cloet:2007em}. Another approach free of these difficulties consists of describing scalar and axial-vector
diquark correlations by a relativistic Faddeev equation in the Nambu Jona-Lasinio
model~\cite{Matevosyan:2011vj}.
\\
(c)
In the $\chi$QSM quarks are not free but bound by a relativistic chiral mean field (semi-classical approximation). This chiral mean field creates a discrete level in the one-quark spectrum
and distorts at the same time the Dirac sea.
The nucleon is obtained by occupying the bound
state with $N_c$ quarks and by filling the negative-continuum Dirac sea. This leads to a Òtwo-componentÓ
structure of nucleon TMDs with distinct contributions from the ÒvalenceÓ bound state level and from
the Dirac sea. Notice that the terms ÒvalenceÓ and ÒseaÓ have a dynamical meaning in the context of
this model, which is not equivalent to the usual meaning of ÒvalenceÓ distributions as the difference
$q-\bar q$ and of ÒseaÓ as $\bar q$. 
The  $\chi$QSM has been applied to describe unpolarized~\cite{Wakamatsu:2009fn,Schweitzer:2012hh} and helicity~\cite{Schweitzer:2012hh} TMDs. It
 predicts significantly  broader $k_\perp$-distributions  for sea-quark TMDs than for valence quarks, 
 due to non-perturbative short-range correlations caused by chiral symmetry breaking effects~\cite{Schweitzer:2012hh}.
A light-front version of the  $\chi$QSM, restricted to the three-quark light-front Fock state, has been also employed to predict all T-even leading-twist TMDs~\cite{Lorce:2011dv} and unpolarized twist-3 TMDs~\cite{Lorce:2014hxa}.

(d) 
The MIT bag model was originally constructed to build the properties of confinement and asymptotic freedom into a simple phenomenological model.
In its simplest version, relativistic non-interacting quarks are confined inside a spherical cavity, with the boundary condition that the quark vector current vanish on the boundary. 
 The assumption of non-interacting quarks is justified by appealing to the idea of asymptotic freedom, whereas the hard boundary condition mimics the binding effects of the gluon fields.
 All leading and sub-leading T-even TMDs have been studied in this model \cite{Avakian:2010br}. The calculation is usually performed using canonical quantization  and expressing the quark field in terms of the   bound-state wave-function obtained from the solution the Dirac equation in the confining potential. 
Alternatively, one could use the formalism of LFWFs in eq. \eqref{phi-overlap3}, by relating the instant-form wave-function to the LFWF as outlined in ref.~\cite{Lorce:2011zta}.
 
(e) The covariant parton model describes the target system as a gas of quasi free partons, i.e. the partons are bound inside the target and behave as free on-mass shell particle at the interaction with the external probe. 
The partons are described by two covariant 3D momentum distributions (one for unpolarized quarks and one for polarized quarks),
which are spherically symmetric in the nucleon rest frame.
This symmetry ultimately connects the distributions of transverse and longitudinal momenta. 
As a consequence, it is possible predict the $x$-and $k_\perp$-dependence of TMDs from the $x-$dependence of known PDFs. 
The model yields results which refer to a large scale.
Other parton model approaches in the context of TMDs were discussed in~\cite{D'Alesio:2009kv,Bourrely:2010ng}.
  
(f) 
In quark-target models the nucleon target is replaced by a dressed quark, consisting of
bare states of a quark and a quark plus a gluon. These models are strictly speaking not nucleon models, but are based on a QCD perturbative description of the quark, including gauge-field degrees of freedom, and therefore provide baseline calculations.

In QCD, the eight TMDs are in principle independent. However, in a large class of quark models
there appear relations among different
TMDs.
 At twist-two level, there
are three flavor-independent relations, two are linear and one is quadratic in the TMDs
 \begin{eqnarray}
 g_{1L}-h_1+\frac{\uvec k_{\perp}^2}{2M^2} h_{1T}^{\perp}&=&0,\label{rel1-tmd}\\
 g_{1T}-h_1+h_{1L}^{\perp}&=&0,\label{rel2-tmd}\\
 (g_{1T})^2+2h_1h_{1T}^{\perp}&=&0\label{rel3-tmd}.
 \end{eqnarray}
Other expressions can be found in the literature, but are just combinations of the relations \eqref{rel1-tmd}-\eqref{rel3-tmd}.
A further flavor-dependent relation involves both polarized and unpolarized TMDs
\begin{eqnarray}
{\cal D}^q f_{1}^{q}+g_{1L}^{q}
&=& 2h_{1}^{q},\label{rel4-tmd}
\end{eqnarray}
where, for a proton target, the flavor factors with $q = u, d$ are given by ${\cal D}^u = 2/3$
and ${\cal D}^d=-1/3$.
These relations were observed in the bag model~\cite{Avakian:2008dz,Avakian:2010br}, in a LFCQM \cite{Pasquini:2008ax}, some quark-diquark models~\cite{She:2009jq,Zhu:2011zza,Jakob:1997wg}, the covariant parton model~\cite{Efremov:2009ze,Efremov:2010mt} and
in the light-front version of the $\chi$QSM \cite{Lorce:2011dv}. Note however
that there also exist models where the relations are not satisfied, like in some versions of the
spectator model~\cite{Bacchetta:2008af} and the quark-target model~\cite{Meissner:2007rx}.
\\
The model relations~\eqref{rel1-tmd}-\eqref{rel4-tmd} are not expected to hold identically
in QCD. They can not be obtained 
 at the level of the operators entering the definition of the TMDs in the correlator~\eqref{corrTMD}, but follow after the calculation of the matrix elements 
 in~\eqref{corrTMD} within a certain model for the nucleon state.
 Furthermore, the TMDs in these relations follow different evolution patterns:
even if the relations are satisfied at some (low) scale, they would not hold anymore for
other (higher) scales. 
Nevertheless, these relations could
still be valid approximations at the threshold between the nonperturbative and perturbative regimes. 
As such, they could be useful conditions to guide TMD parametrizations, providing simplified and intuitive notions for the interpretation
of the data. 
 Note also that some preliminary calculations in lattice QCD give indications that
the relation~\eqref{rel2-tmd} may indeed be approximately satisfied~\cite{Musch:2010ka,Hagler:2009mb}.
As discussed in ref.~\cite{Lorce:2011zta}, it is possible to trace back the origin of the relations~\eqref{rel1-tmd}-\eqref{rel4-tmd} to some 
common simplifying assumptions in the models.
It turns out that  the flavor independent relations~\eqref{rel1-tmd}-\eqref{rel3-tmd} have essentially
a geometrical origin, as was already suspected in the context of the bag model almost a
decade ago~\cite{Efremov:2002qh}. 
Indeed the conditions for their validity are:
1)  the probed quark behaves as if it does not interact directly with the other partons
(i.e. one works within the standard impulse approximation) and there are no explicit
gluons;
2) the quark light-front and canonical polarizations are related by a rotation with axis
orthogonal to both light-cone and $\uvec k_\perp$ directions;
3) the target has spherical symmetry in the canonical-spin basis.
The spherical symmetry is a sufficient but not necessary
condition for the validity of the individual flavor-independent relations \cite{Lorce:2011zta}.  As a matter of fact, a subset of relations can be derived using less restrictive conditions,
like axial symmetry about a specific direction.
For the flavor-dependent relation~\eqref{rel4-tmd}, we need a further condition for the spin-flavor dependent
part of the nucleon wave function, {\it i.e.} 
4) SU(6) spin-flavor symmetry of the nucleon state.

As far as the T-odd TMDs is concerned,  their unique feature 
is the final/initial state interaction effects, which emerge from the gauge-link structure of the parton correlation functions \eqref{corrTMD}.
 Without these
effects, the T-odd parton distributions would vanish.
The first calculation which explicitly predicted a non-zero Sivers function within a scalar-diquark
model~\cite{Brodsky:2002cx} was followed by studies of T-odd TMDs in spectator models~\cite{Gamberg:2003ey,Gamberg:2007wm,Bacchetta:2008af,Goeke:2006ef,Lu:2004au,Lu:2006kt,Ellis:2008in}, bag models \cite{Yuan:2003wk,Courtoy:2008dn,Courtoy:2009pc}, non-relativistic models \cite{Courtoy:2008vi} and a LFCQM \cite{Pasquini:2010af,Pasquini:2011tk}.
In all these model calculation, the  initial/final state interactions
are calculated by taking into account only the leading contribution due to the one-gluon exchange mechanism.
In this way, 
the final expressions for T-odd functions are proportional
to the strong coupling constant, which depends on the intrinsic hadronic scale of the
model. 
A complementary approach is to incorporate 
 the rescattering effects in
augmented LFWFs \cite{Brodsky:2010vs}, containing an imaginary (process-dependent) phase. 
 Further  studies  going beyond the one-gluon exchange approximation
 made used of a model-dependent relation between TMDs and GPDs.
 In this approach, the T-odd TMDs are described via factorization of the  effects of FSIs, incorporated in a so-called "chromodynamics lensing function", and a spatial distortion of impact parameter space parton distributions \cite{Burkardt:2002ks,Burkardt:2003uw}, as discussed in sect.~\ref{sec:6}.
 The interesting new aspect in the work of refs.~\cite{Gamberg:2007wm,Gamberg:2009uk} is the calculation of the lensing function using non-perturbative eikonal methods, which  take into account higher-order gluonic contributions from the gauge link. 

\section{Generalized Parton Distributions }
\label{sec:5}

When integrating the correlator \eqref{eq:GTMD} over $\uvec k_\perp$, one obtains the following quark-quark correlator, denoted as $F$
~\cite{Ji:1996ek,Dittes:1988xz,Mueller:1998fv,Radyushkin:1996nd,Ji:1996nm,Radyushkin:1996ru}

\begin{eqnarray}
&&F^{[\Gamma]}_{\Lambda'\Lambda}(P,x,\Delta,{\cal W}_{straight})=\int\ud^2k_\perp\,W^{[\Gamma]}_{\Lambda'\Lambda}(P,x,\uvec k_\perp,\Delta, {\cal W})\nonumber\\
&&=\int\frac{\ud z^-}{4\pi}e^{ixP^+z^-}\langle p',\Lambda'|\overline\psi(-\tfrac{z}{2})\Gamma\mathcal W\psi(\tfrac{z}{2})|p,\Lambda\rangle\Big|_{z^+=z_\perp=0}.\nonumber\\
&&
\end{eqnarray}
Note that after integration over $\uvec k_\perp$, we are left with a Wilson line connecting directly the points $-\tfrac{z}{2}$ and $\tfrac{z}{2}$ by a straight line. 
The correlator $F$ is parametrized by GPDs at leading twist in the following way
\begin{equation}\label{GPDs}
F^{\mu\nu}=
\left(
\begin{array}{cccc}
\mathcal H&i\frac{\Delta_y}{2M}\mathcal E_T&-i\frac{\Delta_x}{2M}\mathcal E_T&0\\
i\frac{\Delta_y}{2M}\mathcal E&\mathcal H_T+\frac{\Delta^2_x-\Delta^2_y}{2M^2}\tilde{\mathcal H}_T&\frac{\Delta_x\Delta_y}{M^2}\tilde{\mathcal H}_T&\frac{\Delta_x}{2M}\tilde{\mathcal E}\\
-i\frac{\Delta_x}{2M}\mathcal E&\frac{\Delta_x\Delta_y}{M^2}\tilde{\mathcal H}_T&\mathcal H_T-\frac{\Delta^2_x-\Delta^2_y}{2M^2}\tilde{\mathcal H}_T&\frac{\Delta_y}{2M}\tilde{\mathcal E}\\
0&\frac{\Delta_x}{2M}\tilde{\mathcal E}_T&\frac{\Delta_y}{2M}\tilde{\mathcal E}_T&\tilde{\mathcal H}\end{array}\right),
\end{equation}
where we used the following combinations of GPDs
\begin{eqnarray}
&&\mathcal H=\sqrt{1-\xi^2}\left(H-\frac{\xi^2}{1-\xi^2}\,E\right),\nonumber \\
&&\mathcal E=\frac{E}{\sqrt{1-\xi^2}},\nonumber \\
&&\tilde{\mathcal H}=\sqrt{1-\xi^2}\left(\tilde H-\frac{\xi^2}{1-\xi^2}\,\tilde E\right),\nonumber \\
&&\tilde{\mathcal E}=\frac{\xi\,\tilde E}{\sqrt{1-\xi^2}},\nonumber \\
&&\mathcal H_T=\sqrt{1-\xi^2}\left[H_T-\frac{\xi^2}{1-\xi^2}\,E_T+\frac{\frac{\uvec\Delta^2_\perp}{4M^2}\,\tilde H_T+\xi\,\tilde E_T}{1-\xi^2}\right],\nonumber \\
&&\mathcal E_T=\frac{2\tilde H_T+E_T-\xi\,\tilde E_T}{\sqrt{1-\xi^2}},\nonumber \\
&&\tilde{\mathcal H}_T=-\frac{\tilde H_T}{2\sqrt{1-\xi^2}},\nonumber \\
&&\tilde{\mathcal E}_T=\frac{\tilde E_T-\xi\,E_T}{\sqrt{1-\xi^2}}.
\end{eqnarray}
The different entries of \eqref{GPDs} corresponds to different polarization states of the quark and the nucleon, as explained in sect.~\ref{sec:4}. 
Comparing eq.~\eqref{GPDs} with eq.~\eqref{TMDs}, we  notice
 that the same multipole pattern appears, where the role of $\uvec k_\perp$ in eq.~\eqref{TMDs} is played by $\uvec \Delta_\perp$ in eq.~\eqref{GPDs}. 
 
Although direct links between GPDs and TMDs can not exist~\cite{Meissner:2009ww,Meissner:2007rx}, this correspondence leads to expect correlations  between signs or similar orders of magnitude from dynamical origin  \cite{Burkardt:2003uw,Burkardt:2003je}, as we will discuss in sect.~\ref{sec:6}.
\\
The GPDs are function of $x, \, \xi, $ and $ t=-\Delta^2$. They have support in the interval $x\in [-1,1]$, which falls into three regions with different partonic interpretation.
Assuming $\xi>0$,  one has the so-called DGLAP regions with $\xi<x<1$ and $-1<x<-\xi$, where the GPDs describe the emission and reabsorption of a quark and an antiquark pair, respectively,
and the ERBL region with $|x|<\xi$, where the GPDs give the  amplitude for taking out a quark and antiquark from the initial nucleon.
The LFWF overlap representation of the GPDs in the DGLAP region for the quark contribution at $x>\xi$ can be obtained by integrating eq.~\eqref{overlap} over $\uvec k_\perp$, and can be analogously derived  in the other two regions~\cite{Diehl:2000xz,Brodsky:2000xy}.
The LFWF representation makes explicit that a GPD no longer represents a squared amplitude (and
thus a probability), but rather the interference between amplitudes describing different quantum
fluctuations of a nucleon. 
\\
To model GPDs, different approaches are used (for reviews, see~\cite{Goeke:2001tz,Diehl:2003ny,Ji:2004gf,Belitsky:2005qn,Boffi:2007yc,Guidal:2013rya} and reference therein).
One is based on ans\"atze to parametrize GPDs to be applied in phenomenological analyses.
Here, the most popular choice is to parametrize
the hadronic matrix elements which define GPDs in terms of double distributions~\cite{Radyushkin:1997ki,Radyushkin:1998bz}.
\\
Another promising tool in the description of DVCS data makes use of dispersive techniques, which allow one to put additional constraints on GPD parametrizations by relating different observables~\cite{Anikin:2007yh,Radyushkin:2011dh,Kumericki:2007sa,Diehl:2007jb,Goldstein:2009ks,Pasquini:2014vua}.
Other phenomenological applications are based 
on the relation of GPDs to usual parton densities  and form
factors~\cite{Guidal:2004nd,Diehl:2004cx,Diehl:2013xca}.
\\
A different approach consists in a direct calculation of GPDs using effective quark models. After the
first calculation within the MIT bag model~\cite{Ji:1997gm}, GPDs were calculated in the $\chi$QSM \cite{Petrov:1998kf,Penttinen:1999th,Schweitzer:2002nm,Schweitzer:2003ms,Ossmann:2004bp,Wakamatsu:2005vk,Wakamatsu:2008ki}, using covariant Bethe-Salpeter approaches~\cite{Choi:2001fc,Choi:2002ic,Tiburzi:2001je,Tiburzi:2002sw,Tiburzi:2002sx,Theussl:2002xp,Frederico:2009fk,Mezrag:2014jka},
in the quark-target model \cite{Mukherjee:2002pq,Mukherjee:2002xi,Meissner:2007rx}, in nonrelativistic quark models\cite{Scopetta:2002xq,Scopetta:2003et,Scopetta:2004wt}, in light-front quark models \cite{Boffi:2002yy,Boffi:2003yj,Pasquini:2007xz,Pasquini:2006iv,Pasquini:2005dk,Lorce:2011dv}, in meson-cloud models \cite{Pasquini:2004gc,Pasquini:2006dv}, and from LFWFs inspired by AdS/QCD \cite{Vega:2010ns,Vega:2012iz,Chakrabarti:2013gra}.
Although these models single out only a few features of 
 the QCD dynamics in hadrons,
they may provide some guidance on the functional dependence of GPDs, particularly if a model is
constrained to reproduce electromagnetic form factors and ordinary parton distributions.
A more phenomenological approach is to parameterize the LFWFs, and fit the parameters to various nucleon observables. 
In this context, the recent work \cite{Muller:2014tqa} provides an useful start up for phenomenology with LFWFs, using  effective two-body LFWFs within a diquark picture which are consistent with 
Lorentz symmetry requirements and can be used to calculate and 
 correlate different hadronic observables.

\begin{figure}[t]
\resizebox{0.5\textwidth}{!}{
\includegraphics[width=0.5\textwidth]{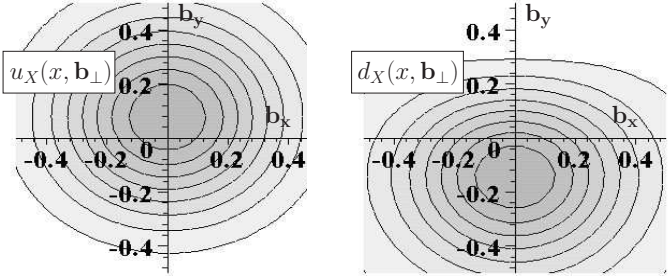}
}
\caption{$q_X$ distribution  of
$u$ and $d$ quarks in the
$\perp$ plane at fixed $x=0.3$ for a proton 
polarized
in the $\hat x$ direction, in the model from
ref. \cite{Burkardt:2002hr}.
For other values of $x$ the deformation looks similar. The signs of
the deformation are determined by the signs of the contribution
from each quark flavor to the proton anomalous magnetic moment.
}
\label{fig:distort}
\end{figure}

 As GTMDs and  Wigner distributions are connected by  Fourier transform at $\xi=0$, one also finds
a direct connection between GPDs at $\xi=0$ and parton distributions in impact-parameter space.  
The multipole structure in $\uvec b_\perp$ is the same as in eq. \eqref{GPDs}, 
with the only difference that there are no polarization effects
in the impact-parameter distributions for longitudinally
polarized quarks in a transversely polarized proton and
vice versa, since they are forbidden by
 time reversal~\cite{Diehl:2005jf}.
At $\xi=0$,  one can also recover a probabilistic interpretation for the impact-parame\-ter space distributions.
For example, the distribution
	of unpolarized quarks in an unpolarized or longitudinally polarized hadron reads \cite{Burkardt:2000za}
\begin{equation}
q(x,{\bf b}_\perp)=\int \frac{d^2\Delta_\perp}{(2\pi)^2} e^{-i{\bf \Delta}_\perp \cdot {\bf b}_\perp}H(x,0,-{\bf \Delta}_\perp^2),
\end{equation}
which corresponds to a cylindrically symmetric distribution.
For a transversely polarized target the transverse polarization singles out a direction and quark distribution
no longer need to be axially symmetric (see fig.~\ref{fig:distort}). The details of the transverse deformation are described by the
helicity flip GPD $E$ (target polarized in $+\hat{x}$ direction) \cite{Burkardt:2002hr}
\begin{eqnarray}
q_{X}(x,{\bf b}_\perp) &=&
\int \frac{d^2\Delta_\perp}{(2\pi)^2} e^{-i{\bf \Delta}_\perp \cdot {\bf b}_\perp}H(x,0,-{\bf \Delta}_\perp^2)
\nonumber\\
&- &\frac{1}{2M} \frac{\partial}{\partial b_y}
\int \frac{d^2\Delta_\perp}{(2\pi)^2} e^{-i{\bf \Delta}_\perp \cdot {\bf b}_\perp}E(x,0,-{\bf \Delta}_\perp^2).\nonumber\\
&&
\label{eq:qUx}
\end{eqnarray}
While details of the GPD $E$ are not known, its $x$-integral is equal to the Pauli form factor $F_2$, which allows one to
constrain the average deformation model-independently to the contribution from each quark flavor $\kappa_q$ to the nucleon anomalous magnetic moment $\kappa^p = \frac{2}{3}\kappa_{u/p}-\frac{1}{3}\kappa_{d/p}$ as
\begin{equation}
\int_0^1 dx \int d^2b_\perp b_y \, q_{X}(x,{\bf b}_\perp) = \frac{\kappa_q}{2M}.
\end{equation}
A similar deformation is observed for quarks with a given transversity in an unpolarized target and described
by a combination of chirally odd GPDs $\bar{E}_T\equiv E_T+2\bar{H}_T$ \cite{Diehl:2005jf}.
In sect. \ref{sec:6} we will discuss the relevance of these deformations for naive T-odd TMDs.

Even before 3D imaging was introduced, there was great interest in GPDs due to their connection with the form
factors of the energy-momentum tensor and thus to the angular momentum (spin plus orbital) carried by quarks of
flavor $q$ as described by the famous Ji relation \cite{Ji:1996ek}
\begin{equation}
J_q = \frac{1}{2}\int_0^1 dx\,x\left[H_q(x,\xi,0)+E_q(x,\xi,0)\right], \label{eq:Ji}
\end{equation}
where the $\xi$-dependence on the r.h.s. disappears upon $x$-integration.

To provide an intuitive derivation for this important result, we first note that for a wave packet $\psi$ describing a nucleon 
polarized in the $+\hat{x}$ direction that is at rest one can use rotational symmetry to eliminate one of the two
terms in the angular momentum
\begin{eqnarray}
J_q^x &=&\int d^3r \langle \psi| yT_q^{0z}({\uvec r})-zT_q^{0y}({\uvec r})|\psi\rangle \\
&=& 2\int d^3r \langle \psi| yT_q^{0z}({\uvec r})|\psi\rangle, \nonumber
\end{eqnarray}
as the two terms on the r.h.s. are up to a sign the same due to rotational invariance about the $\hat{x}$ axis. Here
$T_q^{\mu\nu}$ is the Belifante energy-momentum tensor.
This step is important since GPDs embody information about the distribution of longitudinal momentum in the transverse plane but not {\sl vice versa}. Second we note that we can replace 
\begin{equation}
T_q^{0z}\rightarrow \frac{1}{2}\left[T_q^{00}+
T_q^{0z}+T_q^{z0}+T_q^{zz}\right]=T_q^{++}
\label{eq:T++}
\end{equation} 
in the above integral
since the $2^{nd}$ and $3^{rd}$ term from eq. \eqref{eq:T++} are equal by symmetry of $T^{\mu \nu}$ and
the $1^{st}$ and $4^{th}$ integrate to zero, as both change sign upon rotations by $180^o$ around the $\hat{x}$-axis
due to the presence of the factor $y$ in the integrand,
and thus
\begin{equation}
J_q^x =\int d^3r \langle \psi| yT_q^{++}({\uvec r})|\psi\rangle.
\end{equation}
This result suggests that one can relate the transverse deformation of the impact-parameter dependent parton distributions
(\ref{eq:qUx}) to $J_q^x$. However, this provides only half the answer (the term involving the GPD $E_q$) as
\begin{equation}
q_X({\uvec b}_\perp) \equiv {P^+}^2 \int dx\,x q_X(x,{\bf b}_\perp) =  \langle T^{++}(0,{\bf b}_\perp)\rangle_X
\label{eq:rho++}
\end{equation}
represents the quark distribution 
in impact-parameter space for a transversely localized nucleon - 
not a nucleon at rest. However, since impact-parameter densities are diagonal in the transverse position of the nucleon,
the distribution of quarks in a delocalized wave packet can be obtained through a convolution of the distribution
of the nucleon in the wave packet with the distribution of quarks relative to the center of momentum of the nucleon.
Naively, one may expect that this does not have an effect for a large wave packet of size $R$ as the momenta in a large wave packet are all of ${\cal O}(R)$. However, for the angular momentum the momentum gets multiplied by a factor
$y$ which in the wave packet is ${\cal O}(R)$ and the product is finite in the $R\rightarrow \infty$ limit. 
Indeed, explicit calculations with various wave packets centered at the origin confirm that a large wave packet for a Dirac wave function has its
transverse center of momentum not at the origin but shifted sideways by half a Compton wavelength!
The effect comes from the lower component of the Dirac wavefunction which has an orbital angular momentum that is 
positively correlated to the overall angular momentum of the state and hence the momentum density $T^{0z}$ is
positive for positive $y$ and negative for negative $y$. It is this overall shift which gives rise to the contribution
involving the momentum fraction carried by the quarks in Ji's formula.
With hindsight, it has to be that way because for the Ji relation for a point-like Dirac particle to be satisfied the only
contribution can come from $H(x,0,0)=\delta(x-1)$. Summing up the contribution from the shift of the center of
momentum of the quarks relative to the center of momentum of the nucleon and the overall shift of the center of momentum of the nucleon one thus arrives at eq. (\ref{eq:Ji})~\cite{Burkardt:2005hp}.
\\
\begin{figure}[t]
\resizebox{0.35\textwidth}{!}{
\includegraphics[width=0.35\textwidth]{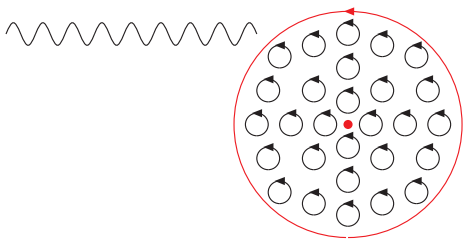}
}
\caption{Quark state with angular momentum out of the plane resulting in counterclockwise quark currents. The oncoming virtual photon 'sees' an enhancement of the
quark distribution above the center of the nucleon.
}
\label{fig:vortex}
\end{figure}
A very similar effect plays a crucial role for
 correlations between the quark distribution in impact-parameter space and quark transversity in a spin-0 or unpolarized target \cite{Diehl:2005jf,Burkardt:2005hp,Burkardt:2007xm}.
As explained above, the Dirac wave function of a quark moving in a spherically symmetric potential (such as in the bag model), with total angular momentum
in the $+\hat{x}$ direction, has its center of momentum shifted in the $+\hat{y}$ direction. For an unpolarized target,
quarks can be polarized in any directions but by rotational invariance there should be a correlation between 
total angular momentum and shift of the center of momentum. More specifically the correlation is such that there is an enhancement
of the quark distribution on the side of the nucleon where quark currents move towards the virtual photon and a corresponding suppression on the opposite side:
An incoming virtual photon
hitting such a quark state with polarization up, quark currents to the left of the center of the bag move towards the virtual photon while those on the right move away from it. Since the relevant quark density in the Bjorken limit involves a $\gamma^+=\frac{1}{\sqrt{2}}\left(\gamma^0+\gamma^3\right)$ matrix structure, this implies that the virtual 
photon in this case 'sees' an enhancement of quarks with spin up on the left and a suppression on the right side of the bag. Similarly, the virtual photon sees an enhancement of quarks with spin to the right above the center of the bag and with spin to the left below the center resulting in a circular polarization distribution as illustrated in fig. \ref{fig:vortex}. 
It turns out that such a pattern is not just a peculiarity of the bag model but can be found in  other quark models that include relativistic effects as the effect arises from the spin-orbit correlation in the
lower component of a relativistic ground state wave function, where the upper component is an s-wave. The same correlation pattern is also found in spectator models 
 and thus appears to be some universal feature for valence quarks in ground state hadrons \cite{Burkardt:2007xm}.
 The overarching principle that explains  the transverse shift of transversely polarized quarks is angular momentum and the Ji relation: for example for a transversely polarized point-like spin $\frac{1}{2}$ particle
at rest, described by a spherically symmetric wave packet, the center of (light-cone) momentum must be shifted by $\frac{1}{2}$ a Compton wavelength for the Ji relation to be satisfied \cite{Burkardt:2007xm,Burkardt:2005hp}. The direction of the shift is sideways relative to its spin.
Lattice QCD calculations of distributions for transversely polarized quarks in   
unpolarized target confirmed the universality of such a pattern by studying quark spin distributions in nucleons as well as pions \cite{Hagler:2009ni} (see fig.~\ref{fig8}).

\section{Chromodynamic Lensing}
\label{sec:6}
As we illustrated in the previous section,
in a target that is polarized transversely ({\it e.g.} vertically), 
the quarks in the target 
can exhibit a significant (left/right) asymmetry of the distribution 
$q_X(x,{\bf b}_\perp)$. In combination with the FSIs, 
this provides a qualitative model for transverse single spin asymmetries, such as the Sivers or Boer-Mulders effects. 

For example, in semi-inclusive DIS, when the 
virtual photon strikes a $u$ quark in a $\perp$ polarized proton,
the $u$ quark distribution is enhanced on the left side of the target
(for a proton with spin pointing up when viewed from the virtual 
photon perspective). 
\\
Although in general the final state 
interaction  is very complicated, we expect it to be on average attractive thus translating a position space
distortion to the left into a momentum space enhancement to the right
and vice versa (fig. \ref{fig:deflect}) \cite{Burkardt:2002ks,Burkardt:2003yg}.
Since this picture is very intuitive, a few words
of caution are in order. First of all, such a reasoning is strictly 
valid only in mean field models for the FSIs as well as in simple
spectator models \cite{Gamberg:2003ey,Gamberg:2007wm,Bacchetta:2008af,Burkardt:2003je,Brodsky:2002rv,Boer:2002ju}. 
Furthermore, even in such mean field or spectator models
there is in general
no one-to-one correspondence between quark distributions
in impact-parameter space and unintegrated parton densities
(e.g. Sivers function) (for a recent overview, see ref.
\cite{Meissner:2007rx}). While both are connected by an overarching Wigner
distribution, 
they are not Fourier transforms of each other.
Nevertheless, since the primordial momentum distribution of the quarks
(without FSIs) must be symmetric, we find a qualitative connection
between the primordial position space asymmetry and the
momentum space asymmetry due to the FSIs.
\begin{figure}[t]
\resizebox{0.5\textwidth}{!}{
\includegraphics[width=0.5\textwidth]{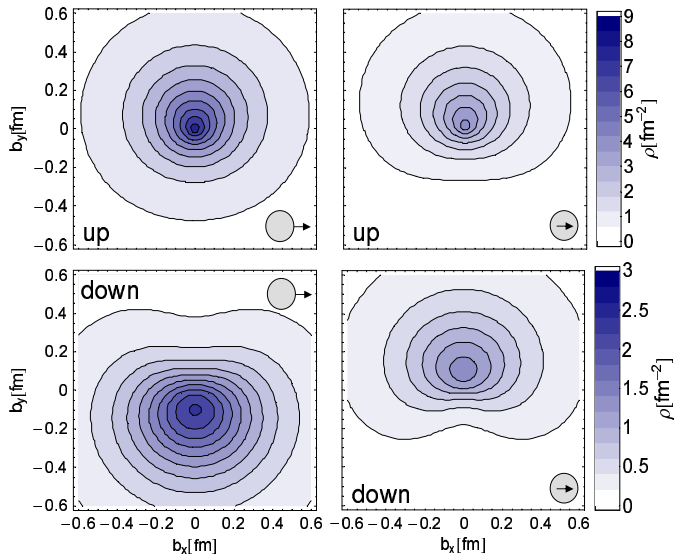}
}
\caption{(Color online) Lattice results~\cite{Gockeler:2006zu} of impact-parameter densities of unpolarized quarks in a transversely polarized nucleon (left) and transversely polarized quarks in an unpolarized nucleon (right) for up (upper panels) and down (lower panels) quarks. The quark spins (inner arrow) and nucleon spins (outer arrow) are oriented in the transverse plane as indicated.}
\label{fig8} \vspace{-0mm}
\end{figure}
Another issue concerns the $x$-dependence of the Sivers function.
The $x$-dependence of the position space asymmetry is described
by the GPD $E(x,0,-\uvec{\Delta}_\perp^2)$. Therefore, within the above
mechanism, the $x$ dependence of the Sivers function should be
related to 
that of $E(x,0,-\uvec{\Delta}_\perp^2)$.
However, the $x$ dependence of $E$ is not known yet and we only
know the Pauli form factor $F_2=\int {\rm d}x E$. Nevertheless, 
if one makes
the additional assumption that $E$ does not fluctuate as a function 
of $x$ then the contribution from each quark flavor $q$ to the
anomalous magnetic moment $\kappa$ determines the sign of 
$E^q(x,0,0)$
and hence of the Sivers function. With these assumptions,
as well as the very plausible assumption that the FSI is on average
attractive, 
one finds that $f_{1T}^{\perp u}<0$, while 
$f_{1T}^{\perp d}>0$. Both signs have been confirmed by a flavor
analysis based on pions produced in a semi-inclusive DIS experiment on a proton target
by the {\sc Hermes} collaboration \cite{Airapetian:2004tw} and are
consistent with a vanishing isoscalar Sivers function observed
by {\sc Compass} \cite{Martin:2007au} as well as more recent {\sc Compass} proton data \cite{Adolph:2014zba}.
\\
A similar mechanism can be invoked to explain the Boer Mulders effect, where  the transverse  spin of the nucleon is replaced by the transverse spin of the struck quark.
As discussed in sect.~\ref{sec:5}, for transversely polarized quarks in an unpolarized nucleon we observe
 a distortion sideways relative to the (transverse) quark spin. 
Combining the above pattern for the shift in impact-parameter space with an attractive FSI, one expects that, in a semi-inclusive DIS experiment,
quarks with polarization up would be preferentially ejected to the right, corresponding to a negative Boer-Mulders function. This result is again consistent both with the {\sc Hermes} and {\sc Compass} experiments 
\cite{Airapetian:2012yg,Adolph:2014pwc}
and also with recent  lattice QCD calculations \cite{Musch:2011er,Engelhardt:2015ypa}.

\begin{figure}[t]
\resizebox{0.45\textwidth}{!}{
\includegraphics[width=0.45\textwidth]{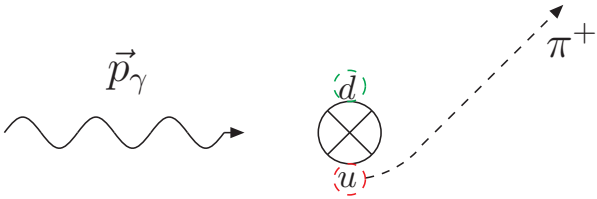}
}
\caption{The transverse distortion of the parton cloud for a proton
that is polarized into the plane, in combination with attractive
FSIs, gives rise to a Sivers effect for $u$ ($d$) quarks with a
$\perp$ momentum that is on the average up (down).}
\label{fig:deflect}
\end{figure}

\section{Higher Twist PDFs}
\label{sec:7}
Higher twist PDFs do not have a simple parton interpretation as number densities or differences of number densities but also involve multi parton correlations. However, some of the correlations still have an intuitive interpretation
as a force \cite{Burkardt:2008ps}.

The longitudinally polarized double-spin asymmetry in DIS in the Bjorken limit is described by the twist-2 PDF
$g_1(x)$ which counts the quarks with spins parallel to the nucleon minus spins anti-parallel.
The twist-3 PDF $g_2(x)$ only enters as a correction in $\frac{1}{Q^2}$. The situation is different for
the double-spin asymmetry for longitudinally polarized leptons and a transversely polarized target, where
$g_1(x)$ and $g_2(x)$ contribute equally. This allows for a clean extraction of $g_2(x)$, which can be written
as 
\begin{equation}
g_2(x)=g_2^{WW}(x)+\bar{g}_2(x),
\end{equation}
where $g_2^{WW}(x)=-g_1(x)+\int_x^1\frac{dy}{y}g_1(y)$
is called Wandzura-Wilczek term. The quark-gluon correlation are contained in $\bar{g}_2(x)$. Using the
QCD equations of motion one finds for example that the first two moments of $\bar{g}_2(x)$ vanish, while
\begin{eqnarray}
d_2&\equiv& 3\int_0^1dx \,x^2\bar{g}_2(x)\\ 
&=&\frac{1}{2M{P^+}^2S^x}
\langle P,S|\bar{\psi}(0)\gamma^+gG^{+x}(0)\psi(0)|P,S\rangle,\nonumber
\end{eqnarray}
i.e. a correlator between the quark density and the same component of the gluon field strength tensor that 
provide the final state interaction force. The only difference to the FSIs in sect. 3 is that the field strength
tensor appears only at the same position as the quark fields, i.e. $d_2$ embodies the FSI force 
\begin{equation}
F_y=-2M^2d_2=-10\frac{\rm GeV}{\rm fm}d_2 \label{eq:Fy}
\end{equation}
for a nucleon polarized in the $+\hat{x}$ direction,
at the
moment immediately after the quark has absorbed the virtual photon. Using the same arguments that were used to
predict the sign of SSA, the signs of the deformation of quark distributions in impact-parameter space can again be
employed to find the direction of the force and hence $d_2$. Furthermore, one can also use the above result to estimate the order of magnitude of $d_2$. A known force is the QCD string tension 
$\sigma \sim 1\frac{\rm GeV}{\rm fm}$. The anticipated force  that results in SSA should be significantly smaller
than this value as the SSA results from an incomplete cancellation of forces in various directions and a value
of $\sim 0.1\frac{\rm GeV}{\rm fm}$ is a reasonable order of magnitude for $|F_y|$. Because of the large prefactor in
\eqref{eq:Fy} this implies a small numerical value $|d_2|\sim 0.01$. The semi-classical connection discussed here between moments of higher-twist PDFs and average transverse forces in DIS perhaps helps to  shed some light on the
fact that the Wandzura-Wilczek approximation for $g_2$ works so well i.e. that $\bar{g}_2(x)$ is so small
\cite{Posik:2014usi,Airapetian:2011wu}. Besides its $x^2$ moment having to be small
to yield a reasonable value for the force, its $x^1$ and $x^0$ moments were already known to be vanishing identically.
Given that its first 3 moments of $\bar{g}_2(x)$ either vanish or are small, it is only reasonable that the whole 
function is not very large.

Similar considerations can be made for the scalar twist 3 PDF $e(x)$. Through a relation similar 
to eq. \eqref{eq:Fy}, its $x^2$ moment can be related to the 'Boer-Mulders force', i.e. to the average transverse
force that acts on a transversely polarized quark in an unpolarized target.

The average transverse force acting on a transversely polarized quark in a longitudinally polarized target has to
vanish due to parity. To be precise, it is the target polarization dependent part that must vanish, as a Boer-Mulders
type force which is independent of the target polarization is still allowed.
The absence of such an average LT-polarization dependent force is consistent with the fact that for the third twist 3 PDF,  i.e. for the twist 3 part of $h_L(x)$, the $x^2$ moment also vanishes identically. However, in this case one finds a simple physical interpretation for its $x^3$ moment as the average longitudinal  gradient of the transverse force acting on a transversely polarized quark in a longitudinally polarized target \cite{Manal}.  

For higher moments than those listed above, simple interpretations have not yet been found.

\section{Conclusions and perspectives}
\label{sec:8}

We reviewed the status of our understanding of nucleon structure based on the modelling of different kinds of parton distributions. 
We started from generalized transverse momentum dependent parton distributions (GTMDs),  which are quark-quark correlators
where the quark fields are taken at the same light-cone time.
They are connected through specific limits or projections to generalized parton distributions (GPDs) and transverse momentum dependent parton distributions (TMDs), accessible in various  semi-inclusive and exclusive processes in the deep inelastic kinematics.
Through a Fourier transform of the GTMDs in the transverse space,  one obtains Wigner distributions
which represent the quantum-mechanical analogues of the classical phase-space distributions. 
All these distributions entail different and complementary information about the quark dynamics inside the nucleon.
Due to the difficulty of solving QCD in the
nonperturbative regime, models are of utmost importance in the study of these distributions.
QCD inspired models often mimic isolated features of the underlying theory, but are crucial for the physical  interpretation of the parton distributions and to gain an intuitive physical  picture that captures the essential mechanisms coming into play in deep inelastic scattering processes.
Specific questions that we addressed in this work  are  about the angular momentum structure of the proton, the role of spin-orbit correlations of quarks in the transverse-coordinate and transverse-momentum space, and the mechanism
of color forces acting on  quarks in deep inelastic scattering processes.
In particular, we applied the Wigner distributions to give a simple interpretation of the difference between the Jaffe-Manohar and Ji orbital angular momentum of the quarks.
We presented an intuitive derivation of the Ji's relation between the generalized parton distributions and the quark angular momentum.
We discussed the role of final/initial state interactions in deep inelastic processes,  
and we summarized
 recent results on relating higher-twist parton distribution functions  to the transverse color forces acting on quarks
in deep-inelastic scattering.
The resulting pictures for the GPDs in impact-parameter space and TMDs in momentum space have been confirmed by available lattice studies~\cite{Hagler:2009ni}. 
However, for a comprehensive spatial and momentum imaging 
 of quark distributions from experimental data, the information  from the present measurements at JLab, HERMES and COMPASS have to be supplemented from the upcoming data from COMPASS~\cite{Gautheron:2010wva},  the 12 GeV JLab program~\cite{Burkert:2012rh} and  the experimental program at an EIC~\cite{Accardi:2012qut,Boer:2011fh}.
 Exploratory lattice QCD studies are under way to estimate not only the Sivers and Boer-Mulders effect, but
 to calculate also the Jaffe-Manohar OAM. 
Finally, upcoming results from JLab experiments~\cite{HallB-ht,HallB-ht2} will provide new information on higher-twist TMDs and GPDs. 
In particular, from the measurements of twist-3 GPDs, it  would be useful to extract experimental constraints on local correlation functions
  that could give useful insights about 
the interpretation of
the difference between the Ji and Jaffe-Manohar OAM in terms of local torque,
as  presented in this work.

\begin{acknowledgement} 
This work was supported in part by DOE  (FG03-95ER40965) and in part from the European Research Council (ERC) under the European UnionÕs Horizon 2020 research and innovation programme (grant agreement No. 647981, 3DSPIN).
\end{acknowledgement}

\end{document}